\documentstyle[aps]{revtex}

\begin{document}
\title{{\bf Anomalies in Ward Identities for Three-Point Functions Revisited}}
\author{O.A. Battistel$^*$ and O.L. Battistel$^{**}$}
\maketitle

\centerline{*Dept. of Physics}

\centerline{Universidade Federal de Santa
Maria, 5093, 97119-900, Santa Maria, RS, Brazil}

\centerline{**Exact Science Area}

\centerline{Centro Universit\'ario Franciscano, Rua dos
Andradas, 1614, 97010-032, Santa Maria, RS, Brazil }

\begin{abstract}
A general calculational method is applied to investigate symmetry relations
among divergent amplitudes in a free fermion model. A very traditional work
on this subject is revisited. A systematic study of one, two and three point
functions associated to scalar, pseudoscalar, vector and axial-vector
densities is performed. The divergent content of the amplitudes are left in
terms of five basic objects (external momentum independent). No specific
assumptions about a regulator is adopted in the calculations. All
ambiguities and symmetry violating terms are shown to be associated with
only three combinations of the basic divergent objects. Our final results
can be mapped in the corresponding Dimensional Regularization calculations
(in cases where this technique could be applied) or in those of Gertsein and
Jackiw which we will show in detail. The results emerging from our general
approach allow us to extract, in a natural way, a set of reasonable
conditions (e.g. crucial for QED consistency) that could lead us to obtain
all Ward Identities satisfied. Consequently, we conclude that the
traditional approach used to justify the famous triangular anomalies in
perturbative calculations could be questionable. An alternative point of
view, dismissed of ambiguities, which lead to a correct description of the
associated phenomenology, is pointed out.
\end{abstract}

\vskip0.3cm

\section{Introduction}

Quantum Field Theory is nowadays our main tool for the investigation of the
elementary particle interactions. This is also due to the fact that it
allows us to study the consequences of symmetries assumed relevant, our
fundamental working hypothesis. Since this is seldom possible, in an exact
way, perturbative techniques become most relevant. Unfortunately, this type
of solution is plagued with mathematical problems coming from divergent
integrals. Therefore we are forced to have the recourse of auxiliary
techniques based on regularization schemes in order to extract the physical
content of calculated amplitudes. However, the regularization schemes
invariably modify the amplitudes, as they are dictated by Feynman rules, and
consequently provide only a particular interpretation for the mathematical
indefinitions in the problem. Frequently, the final results become
regularization-scheme dependent and may not reflect the full dynamical
content of the underlying theory. In this context two of the main issues are
the question of ambiguities associated with arbitrary momentum routing in
loops and the symmetry relation violations \cite{ref1}. Many of these
fundamental questions have been adequately solved after the construction of
Dimensional Regularization (DR) \cite{ref2}. There is however one important
issue which, due to the mathematical limitation of the technique, could not
be handled, namely, the pseudo-amplitudes. Perhaps this is the reason why
even nowadays the treatment given to such amplitudes have the recourse of
methods completely discarded in all cases where DR applies. We refer
specifically to the problem of triangle anomalies, which has been discussed
in the pioneer work of Gertsein and Jackiw \cite{ref1}. They proposed that
the ambiguous character of the amplitudes is associated to the existence of
violations in symmetry relations. In their analysis, the surface terms
associated to eventual shifts relating two different choices for the
internal (arbitrary) momentum routing, play an important role. In spite of
the relative success of this recipe there is a deeply frustrating aspect
involved, namely, the assumption that the space-time homogeneity
(translational invariance), one of the most basic symmetry implemented in
the construction of quantum field theories, is lost in the perturbative
calculations. In addition, the predictive power of such theories are
affected because in cases where the amplitudes assume degrees of divergence
higher than logarithmic this means that no unique answer can be obtained 
\cite{ref3}. To decide by one of the infinite possible expressions for a
physical amplitude we need some knowledge about the phenomenology or, in
other words, to choose a value for each ambiguity involved. But there are
some situations where a completely healthy amplitude cannot be accessed by
choosing ambiguities. This is exactly the situation of the triangle
anomalies \cite{ref4}. This point of view, introduced by Gertsein and Jackiw 
\cite{ref1}, before the advent of DR technique, remains as the accepted one
even in modern quantum field theory text books \cite{ref5}, \cite{ref6}: the
violation of symmetry relations (anomalies) in perturbative calculations is
associated to the divergent (ambiguous) character of the amplitudes. This is
a completely discarded picture in all other situations involving divergences
in perturbative evaluations of physical amplitudes. Anomalies, in other
hand, are shown to emerge even in theories dismissed of divergences \cite
{ref6}, \cite{ref7} and predicted by several arguments in QFT \cite{ref6}, 
\cite{ref8}. It seems to have no doubt about their existence. So, if there
is a fundamental quantum phenomenon of the nature associated to anomalies 
\cite{ref9}, it must be present even in exact solutions of the theory
(dismissed of infinities and ambiguities). Therefore it is reasonable to
believe that it must be possible to establish their existence without having
to appeal to ambiguities, even in perturbative calculations where divergent
amplitudes are involved.

Ambiguities and symmetry violations in perturbative calculations, in other
hand, remain as unsolved questions even for many others than anomaly
problems. Discussions involving these aspects can be copiously found in the
recent literature in different contexts but playing a crucial role in the
job of the QFT theoretical methods description of a particular
phenomenology. In the case of effective non-renormalizable models, like NJL
model \cite{ref10}, ambiguities and symmetry violations are deeply related
with the own predictive power of such models \cite{ref11}. Investigations on
this subject have pointed out that usual procedures like DR, sharp cut off
and Pauli-Villars cannot offer acceptable results \cite{ref12}. More
recently the role of ambiguities have emerged in the context of Lorentz and
CPT violations in an extended version of Standard Model \cite{ref13}.
Perturbative evaluations of the radiative contributions to a Lorentz and CPT
breaking coupling of the pure photon sector of an extended QED, of the
Chern-Simmons form, are verified to be regularization-scheme dependent and
plagued by the ambiguities. In addition many aspects of the discussions
about the controversy involving the Chern-Simmons shift are closely related
to that of $AVV$ anomaly once the mathematical structure of the (divergent)
contributions are completely similar \cite{ref14}. Explicitly evaluation of
the radiatively induced Chern-Simmons term using DR \cite{ref15},
Differential Regularization \cite{ref16} and surface term evaluation \cite
{ref14}, among others \cite{ref17}, cannot lead us to a definite value,
i.e., only an ambiguous value can be extracted using traditional techniques.

Ambiguities, symmetry relation violations and regularization-scheme
dependence are also present in the aim of the investigations about the
implications of the fermionic tensor densities for low energy hadronic
observable, a completely new physics \cite{ref18}.

All the above cited problems, and certainly others omitted here, constitute
strong indications that additional investigations on the subject of
ambiguities, symmetry relation violations and regularization-scheme
dependence in perturbative calculations are in order. In light of these
statements we have devoted this work to the investigation of the ambiguities
and symmetry relations in a free (Dirac) fermion model. This is not more
than an amplified revisiting to this question in a closely related way with
the study that was done by Gertsein and Jackiw in ref.\cite{ref1}, a
historical paper on this subject. Our intention is to use a general
calculational method to manipulate and calculate the divergent integrals in
such a way that all amplitudes in different theories and models are handled
by the same prescription and connections with conventional techniques could
be established. A such alternative strategy to manipulate and calculate
divergences in perturbative calculations have been proposed \cite{ref19} and
successfully applied in the treatment of the ambiguities and symmetry
relation violations in the context of the gauged NJL Model \cite{ref20} and
to investigate the possibility of a radiatively induced Chern-Simmons shift
in an extended version of QED \cite{ref21}. This method so far has proved to
be an adequate tool to treat all questions related with divergences from the
renormalization of standard theories like QED \cite{ref22}, calculation of
renormalization group coefficients \cite{ref23} and, for calculations in
effective non-renormalizable theories \cite{ref19}. Having this in mind we
turn attention in the present work perhaps to the historically most relevant
question that involves the association between the violation of symmetry
relations and ambiguities. The advantage of the method adopted here is that
it allows an immediate connection with other current approaches. In
particular, we can use the consistent results obtained by DR, where it
applies, as element of the analysis for the search of an universal
interpretation for the divergences in physical amplitudes. From the
theoretical point of view an adequate description of the anomaly phenomenon,
and any other in the perturbative calculations, is only achieved if for
identical mathematical objects an identical value is attributed in all
situations where it occurs. In this context, it is most desirable to obtain
the anomalies in a natural way within a context which treats all anomalous
and non-anomalous amplitudes according to the same scheme. In this direction
the first step is the verification if the current traditional approach used
to justify the triangle anomalies produces results in a consistent way with
those of Dimensional Regularization where both procedures could be applied.
One must check whether the current way as followed in ref.\cite{ref1} for
calculating e.g. the famous axial anomaly remains appropriate when used to
investigate other symmetry relations. In other words, anomalies must also
emerge in a treatment designed to consistently handle all other amplitudes,
i.e., where ambiguities are not necessarily present, in spite of the
divergent character.

After detailed results have been obtained, in the point of view of our
general calculational method, we will show that the results can be mapped
separately in those corresponding of Dimensional Regularization and to those
of surface's terms analysis of ref.\cite{ref1}, but these two prescriptions
are conflicting, i.e., cannot be simultaneously mapped by an universal
interpretation. An important aspect involved in this investigation is that
when we demand a complete mapping of our results (in 4-D) with those of DR
we obtain a set of conditions which are not compatible with any symmetry
violations, even for the anomalous cases. This is an indication that the
procedure adopted in ref.\cite{ref1} to establish the triangle anomaly
cannot be considered as a consistent one. We conclude that this is due to
the fact that only two-point functions are explicitly calculated in the
traditional procedure and the violation of symmetry relations in the
anomalous amplitudes are intrinsic properties of the three-point functions
involved. A consistent calculation of the $AVV$ and $AAA$ amplitudes reveals
that the violations of the Ward Identities characterizing the anomaly
phenomenon emerges, in a natural way, with the expected value, without
having to admit ambiguities and within an interpretation which is universal
in a sense that consistent results are obtained for all amplitudes in the
same point of view \cite{ref19}.

As discussed before we consider the free fermion model of ref.\cite{ref1}
but discarding internal symmetries, which are irrelevant for our purposes.
In section II we define the model, notation and relevant Ward Identities. In
section III we briefly establish the operational strategy for the
manipulation and calculation of divergent amplitudes. In section IV a study
of ambiguities is presented in one and two-point functions calculated
explicitly with arbitrary momentum routing. We also explicitate the
ambiguous terms of the three-point functions. In section V we reproduce
(from our results) those of ref.\cite{ref1}. In section VI we discuss Ward
Identities. The final remarks are contained in section VII.

\section{Definitions, Notation and Current Algebra Results for the Model}

We start introducing the notation to be used and defining the quantities we
will be concerned with for the rest of the work. We closely follow ref.\cite
{ref1} with which we compare our results.

Let us consider a spin 1/2, mass $m$ free fermion model. There will be
therefore a massive field which obeys Dirac's equation and with which we can
construct currents $j_{i}(x)$ defined by 
\begin{equation}
j_{i}(x)=\bar{\psi}(x)\Gamma _{i}\psi (x),
\end{equation}
where $\Gamma _{i}$ are the Dirac's matrices responsible for the
transformation properties of the currents: 
\begin{equation}
\Gamma _{i}=[\hat{1};\gamma _{5};\gamma _{\mu };i\gamma _{\mu }\gamma _{5}],
\end{equation}
characterizing the scalar $S(x)$, pseudo-scalar $P(x)$, vector $V_{\mu }(x)$
and axial-vector $A_{\mu }(x)$ densities, respectively. Tensor densities
will be considered elsewhere \cite{Tensor}. An important property in this
model is the value of the four divergences 
\begin{equation}
\left\{ 
\begin{array}{c}
\partial _{\mu }V_{\mu }(x)=0 \\ 
\partial _{\mu }A_{\mu }(x)=2mP(x),
\end{array}
\right.
\end{equation}
and their commutation relations at equal times \cite{ref1}. With the above
definitions and the fermionic propagator 
\begin{equation}
iS_{F}(p)=\frac{i}{\not{p}-m},
\end{equation}
it is possible to construct n-point functions, which we define in the same
way as in ref.\cite{ref1} as follows

\begin{itemize}
\item  One-point functions: 
\begin{equation}
T^{i}(k_{1},m)=\int \frac{d^{4}k}{(2\pi )^{4}}Tr\left\{ \Gamma _{i}\frac{1}{%
[(\not{k}+\not{k}_{1})-m]}\right\} ,
\end{equation}

\item  Two-point functions: 
\begin{equation}
T^{ij}(k_{1},m;k_{2},m)=\int \frac{d^{4}k}{(2\pi )^{4}}Tr\left\{ \Gamma _{i}%
\frac{1}{[(\not{k}+{\not{k}}_{1})-m]}\Gamma _{j}\frac{1}{[(\not{k}+{\not{k}}%
_{2})-m]}\right\} ,
\end{equation}

\item  Three-point functions:
\end{itemize}

\begin{eqnarray}
&&\!\!\!\!\!\!\!\!\!T^{lij}(k_{1},m;k_{2},m;k_{3},m)=  \nonumber \\
&&\int \frac{d^{4}k}{(2\pi )^{4}}Tr\left\{ \Gamma _{l}\frac{1}{[(\not{k}+{%
\not{k}}_{3})-m]}\Gamma _{i}\frac{1}{[(\not{k}+{\not{k}}_{1})-m]}\Gamma _{j}%
\frac{1}{[(\not{k}+{\not{k}}_{2})-m]}\right\} ,
\end{eqnarray}
and so on. Here the $k_{1}$, $k_{2}$ and $k_{3}$ represent the arbitrary
choices for the internal momenta of the loop. Energy momentum conservation
only requires these quantities must be related with the external momenta,
e.g., in the three-point functions we have: $k_{3}-k_{1}=p$, $%
k_{1}-k_{2}=p^{\prime }$ and $k_{3}-k_{2}=q$.

In our notation the vertex to the left is assumed to be connected with the
``initial state'' with external momentum $q$ and the vertex operator $\Gamma
_{l}$. The other two vertices correspond to ``final states'' with external
momenta $p$ and $p^{\prime }$ and the vertex operators $\Gamma _{i}$ e $%
\Gamma _{j}$, respectively. Besides the upper indices are associated, in the
order they appear, with the respective Lorentz indices in the same order,
whenever is the case. As an example, if $\Gamma _{l}=\gamma _{\lambda }$, $%
\Gamma _{i}=i\gamma _{\mu }\gamma _{5}$ and $\gamma _{j}=\gamma _{5}$ we
have 
\begin{equation}
T_{\lambda \mu }^{VAP}(k_{1},m;k_{2},m;k_{3},m).
\end{equation}
This notation emphasizes the masses and momenta carried by the internal
lines and characterizes each diagram completely. In particular this means
that in the case we have one particle in the initial state and two on the
final one, symmetrization in the final states will be required. For example
for the process $S\rightarrow VV$ we define the corresponding amplitude as

\begin{equation}
T^{S\rightarrow VV}_{\mu\nu}=T^{SVV}_{\mu\nu}(k_1,m;k_2,m;k_3,m)+
T^{SVV}_{\nu\mu}(l_1,m;l_2,m;l_3,m).
\end{equation}
The first term represents the direct channel and the second the crossed
channel. In the second term $l_1$, $l_2$ e $l_3$ are the arbitrary choices
for the corresponding internal momenta.

There are integral representations for the functions defined above with the
Fourier transform of the currents 
\begin{equation}
<j_{1}(q)j_{2}(-q)>\equiv \int e^{-ipx}d^{4}x<0|T(j_{1}(x)j_{2}(0)|0>,
\end{equation}
in the case of two-point functions, and 
\begin{equation}
<j_{1}(p)j_{2}(p^{\prime })j_{3}(q)>\equiv \int e^{-ipx}e^{-ip^{\prime
}y}d^{4}xd^{4}y<0|T(j_{1}(x)j_{2}(y)j_{3}(0))|0>,
\end{equation}
for the three-point functions. With these elements and the standard methods
of the current algebra \cite{ref1},\cite{ref4}, one can establish relations
among the n-point functions, i.e., Ward Identities. For the simple model in
question such symmetry relations are reduced to the conservation of the
vector current and the well-known proportionality of the divergence of the
axial current and the pseudoscalar one. It is important to remark that such
relations are \underline{exact} and should be satisfied in spite of the
divergent character of up to $n$-point functions.

This brief discussion summarizes the dilemma one has to face when
calculating divergent amplitudes: to maintain their properties \underline{%
after the calculation}. {\it It is our purpose to investigate under which
conditions it is possible to evaluate these amplitudes and to get consistent
results in what concerns ambiguities and symmetry relations}.

The Ward Identities that we should verify are the following

\begin{itemize}
\item  {\it One-point function}

\begin{equation}
T^V_\mu (k_1,m)=0
\end{equation}

\item  {\it Two-point functions}

\begin{equation}
(k_1-k_2)_\mu T^{VS}_\mu (k_1,m;k_2,m)=0
\end{equation}

\begin{equation}
(k_1-k_2)_\mu T^{VV}_{\mu\nu} (k_1,m;k_2,m)=0
\end{equation}

\begin{equation}
(k_1-k_2)_\nu T^{VV}_{\mu\nu} (k_1,m;k_2,m)=0
\end{equation}

\begin{equation}
(k_1-k_2)_\mu T^{AP}_{\mu} (k_1,m;k_2,m)=-2miT^{PP}(k_1,m;k_2,m)
\end{equation}

\begin{equation}
(k_1-k_2)_\mu T^{AV}_{\mu\nu} (k_1,m;k_2,m)=-2miT^{PV}_{\nu}(k_1,m;k_2,m)
-iT^{S}(k_1,m)-iT^{S}(k_2,m)
\end{equation}

\begin{equation}
(k_1-k_2)_\nu T^{AV}_{\mu\nu} (k_1,m;k_2,m)=0
\end{equation}

\begin{equation}
(k_1-k_2)_\mu T^{AA}_{\mu\nu} (k_1,m;k_2,m)=-2miT^{PA}_{\nu}(k_1,m;k_2,m)
\end{equation}

\begin{equation}
(k_1-k_2)_\nu (k_1-k_2)_\mu T^{AA}_{\mu\nu} (k_1,m;k_2,m)=
(2m)^2T^{PP}(k_1,m;k_2,m)+2mT^{S}(k_1,m)+2mT^{S}(k_2,m)
\end{equation}

\item  {\it Three-point functions} 
\begin{equation}
q_{\lambda }T_{\lambda }^{V\rightarrow SS}=0
\end{equation}

\begin{equation}
q_\lambda T^{V\rightarrow PP}_\lambda =0
\end{equation}

\begin{equation}
q_\lambda T^{A\rightarrow SP}_\lambda =-2miT^{P\rightarrow SP}
\end{equation}

\begin{equation}
p^{\prime}_\nu T^{S\rightarrow VV}_{\mu \nu} =0
\end{equation}

\begin{equation}
p_\mu T^{S\rightarrow VV}_{\mu \nu} =0
\end{equation}

\begin{equation}
p_\mu T^{S\rightarrow AA}_{\mu \nu} =2miT^{S\rightarrow PA}_\nu
\end{equation}

\begin{equation}
p^{\prime}_\nu T^{S\rightarrow AA}_{\mu \nu} =2miT^{S\rightarrow AP}_\mu
\end{equation}

\begin{equation}
p_\mu T^{P\rightarrow AV}_{\mu \nu} =2miT^{P\rightarrow PV}_\nu
\end{equation}

\begin{equation}
p^{\prime}_\nu T^{P\rightarrow AV}_{\mu \nu} =0
\end{equation}

\begin{equation}
p_\mu T^{P\rightarrow VV}_{\mu \nu} =0
\end{equation}

\begin{equation}
p^{\prime}_\nu T^{P\rightarrow VV}_{\mu \nu} =0
\end{equation}

\begin{equation}
p_\mu T^{S\rightarrow AV}_{\mu \nu} =2miT^{S\rightarrow PV}_\nu
\end{equation}

\begin{equation}
p^{\prime}_\nu T^{S\rightarrow AV}_{\mu \nu} =0
\end{equation}

\begin{equation}
p_\mu T^{P\rightarrow AA}_{\mu \nu} =2miT^{P\rightarrow PA}_\nu
\end{equation}

\begin{equation}
p^{\prime}_\nu T^{P\rightarrow AA}_{\mu \nu} =2miT^{P\rightarrow AP}_\mu
\end{equation}

\begin{equation}
q_\lambda T^{A\rightarrow VV}_{\lambda \mu \nu} =-2miT^{P\rightarrow
VV}_{\mu\nu}
\end{equation}

\begin{equation}
p_\mu T^{A\rightarrow VV}_{\lambda \mu \nu} =0
\end{equation}

\begin{equation}
p^{\prime}_\nu T^{A\rightarrow VV}_{\lambda \mu \nu} =0
\end{equation}

\begin{equation}
q_\lambda T^{A\rightarrow AA}_{\lambda \mu \nu} =-2miT^{P\rightarrow
AA}_{\mu\nu}
\end{equation}

\begin{equation}
p_\mu T^{A\rightarrow AA}_{\lambda \mu \nu} =2miT^{A\rightarrow
PA}_{\lambda\nu}
\end{equation}

\begin{equation}
p^{\prime}_\nu T^{A\rightarrow AA}_{\lambda \mu \nu} =-2miT^{A\rightarrow
PA}_{\lambda\mu}
\end{equation}

\begin{equation}
q_\lambda T^{V\rightarrow VV}_{\lambda \mu \nu} =0
\end{equation}

\begin{equation}
p_\mu T^{V\rightarrow VV}_{\lambda \mu \nu} =0
\end{equation}

\begin{equation}
p^{\prime}_\nu T^{V\rightarrow VV}_{\lambda \mu \nu} =0
\end{equation}

\begin{equation}
q_\lambda T^{V\rightarrow AA}_{\lambda \mu \nu} =0
\end{equation}

\begin{equation}
p_\mu T^{V\rightarrow AA}_{\lambda \mu \nu} =2miT^{V\rightarrow
PA}_{\lambda\nu}
\end{equation}

\begin{equation}
p^{\prime}_\nu T^{V\rightarrow AA}_{\lambda \mu \nu} =2miT^{V\rightarrow
AP}_{\lambda\mu}
\end{equation}
\end{itemize}

\section{Calculational Strategy to Manipulate Divergent Integrals}

For the verification of the above cited Ward Identities it becomes necessary
some explicitly evaluation of the divergent amplitudes. In face of the
well-known mathematical indefinitions involved we need to adopt some
strategy to handle this problem. The usual procedure is the adoption of a
regularization scheme or any equivalent philosophy. For the following
calculations we will adopt an alternative strategy. Rather than the
specification of some regularization, to justify all the manipulations, we
will assume the presence of a regulating distribution only in an implicitly
way. Schematically: 
\begin{equation}
\int \frac{d^{4}k}{\left( 2\pi \right) ^{4}}f(k)\rightarrow \int \frac{d^{4}k%
}{\left( 2\pi \right) ^{4}}f(k)\left\{ \lim_{\Lambda _{i}^{2}\rightarrow
\infty }G_{\Lambda _{i}}\left( k^{2},\Lambda _{i}^{2}\right) \right\}
=\int_{\Lambda }\frac{d^{4}k}{\left( 2\pi \right) ^{4}}f(k).
\end{equation}
Here $\Lambda _{i}^{\prime }s$ are parameters of the generic distribution $%
G(\Lambda _{i}^{2},k^{2})$. In addition to the obvious finiteness of the
modified integral, we require two very general properties of $G(\Lambda
_{i}^{2},k^{2})$; it should be even in the integrating momentum $k$, due to
Lorentz invariance maintenance, and a well-defined connection limit should
exists, i.e., $\lim_{\Lambda _{i}^{2}\rightarrow \infty }G_{\Lambda
_{i}}\left( k^{2},\Lambda _{i}^{2}\right) =1$. The first property imply that
all odd integrands vanishes. The second one guarantees, in particular, that
the value of finite integrals in the amplitudes will not be modified. In the
next step we manipulate the integrand of the divergent integrals by using
identities to generate a mathematical expression where all the divergences
are contained in external momenta independent structures. This goal can be
achieved by the use of an adequate identity like: 
\begin{equation}
\frac{1}{[(k+k_{i})^{2}-m^{2}]}=\frac{1}{(k^{2}-m^{2})}+\sum_{j=1}^{N}\frac{%
\left( -1\right) ^{j}\left( k_{i}^{2}+2k_{i}\cdot k\right) ^{j}}{\left(
k^{2}-m^{2}\right) ^{j+1}}+\frac{\left( -1\right) ^{N+1}\left(
k_{i}^{2}+2k_{i}\cdot k\right) ^{N+1}}{\left( k^{2}-m^{2}\right) ^{N+1}\left[
\left( k+k_{i}\right) ^{2}-m^{2}\right] },
\end{equation}
where $k_{i}$ is (in principle) an arbitrary momentum. The minor value of $N$
should be the one that leads the last term in above expression to a finite
integral, and then, in face of the connection limit requirement, the
corresponding integration can be performed without restriction. The
divergent structures obtained this way have no additional assumptions and
(in the discussed problem) they are written as a combination of five
objects, namely: 
\begin{eqnarray}
\bullet \Box _{\alpha \beta \mu \nu } &=&\int_{\Lambda }\frac{d^{4}k}{\left(
2\pi \right) ^{4}}\frac{24k_{\mu }k_{\nu }k_{\alpha }k_{\beta }}{\left(
k^{2}-m^{2}\right) ^{4}}-g_{\alpha \beta }\int_{\Lambda }\frac{d^{4}k}{%
\left( 2\pi \right) ^{4}}\frac{4k_{\mu }k_{\nu }}{\left( k^{2}-m^{2}\right)
^{3}} \\
&&-g_{\alpha \nu }\int_{\Lambda }\frac{d^{4}k}{\left( 2\pi \right) ^{4}}%
\frac{4k_{\beta }k_{\mu }}{\left( k^{2}-m^{2}\right) ^{3}}-g_{\alpha \mu
}\int_{\Lambda }\frac{d^{4}k}{\left( 2\pi \right) ^{4}}\frac{4k_{\beta
}k_{\nu }}{\left( k^{2}-m^{2}\right) ^{3}}  \nonumber \\
\bullet \Delta _{\mu \nu } &=&\int_{\Lambda }\frac{d^{4}k}{\left( 2\pi
\right) ^{4}}\frac{4k_{\mu }k_{\nu }}{\left( k^{2}-m^{2}\right) ^{3}}%
-\int_{\Lambda }\frac{d^{4}k}{\left( 2\pi \right) ^{4}}\frac{g_{\mu \nu }}{%
\left( k^{2}-m^{2}\right) ^{2}} \\
\bullet \nabla _{\mu \nu } &=&\int_{\Lambda }\frac{d^{4}k}{\left( 2\pi
\right) ^{4}}\frac{2k_{\nu }k_{\mu }}{\left( k^{2}-m^{2}\right) ^{2}}%
-\int_{\Lambda }\frac{d^{4}k}{\left( 2\pi \right) ^{4}}\frac{g_{\mu \nu }}{%
\left( k^{2}-m^{2}\right) } \\
\bullet I_{log}(m^{2}) &=&\int_{\Lambda }\frac{d^{4}k}{\left( 2\pi \right)
^{4}}\frac{1}{\left( k^{2}-m^{2}\right) ^{2}} \\
\bullet I_{quad}(m^{2}) &=&\int_{\Lambda }\frac{d^{4}k}{\left( 2\pi \right)
^{4}}\frac{1}{\left( k^{2}-m^{2}\right) }.
\end{eqnarray}
With this strategy, it became possible to map the final expressions obtained
by us into the corresponding results of other techniques, once all the steps
are perfectly valid with reasonable regularization prescriptions, including
those of DR. All we need is to evaluate the divergent structures obtained
with the specific philosophy which we will make contact. In addition {\it we
call attention to the fact that no shifts will be used on the general
routing assumed for all amplitudes. Consequently it will be possible to make
contact with those results corresponding to explicit evaluation of surfaces
terms in eventual shifts performed in the integrating momentum of the loop
integrals} \cite{ref1}. This very general character of the adopted strategy
will become the most important ingredient for the analysis we want to do and
to support our conclusions. Another important aspect of the procedure is
that a definite value for each divergent integral involved is attributed.
This value is used always that the integrals are present in a physical
amplitude in all theories and models, providing, in this way, an universal
point of view for the problem. No expansions, limits or not totally
controlled procedures are included. All the manipulations and calculations
we need in treatment of divergences in QFT is performed without the explicit
calculation of a divergent integral as become clear in what follows.

\section{Ambiguities}

The integral representation, eq.(5), for one-point functions indicates, by
power counting, that there is a cubic degree of divergence. Also, eq.(6)
reveals that the two-point functions are quadratically divergent and eq.(7)
indicates a linearly one for the three-point functions. In the point of view
adopted in ref.\cite{ref1}, all these quantities are, in principle,
ambiguous quantities. The reason for this resides on the fact that the
energy-momentum conservation relations do not uniquely specify the internal
momenta in the loop. It is possible to make different choices for the
internal momentum label. Such choices could only be equivalent if shifts in
the integrating variable were allowed, which is not the case when the degree
of divergence is higher than logarithmic \cite{ref5}, \cite{ref6}. In what
follows we will explicitly evaluate one and two-point functions in detail
from the point of view of our general calculational method. For the
three-point functions only the potentially ambiguous terms are explicitly
showed as required for the analysis we want to do, closely related to that
of ref.\cite{ref1}. All calculations will be made adopting general routing
for the internal momenta in such a way all possible choices are
automatically included. After that all desired results have been obtained
and a mapping to those of ref.\cite{ref1} produced, we will analyze all
possible ambiguities. An amazing simple structure for these ambiguities will
emerge so that immediate and transparent conclusions can be extracted.

\subsection{One-Point Functions}

We start by considering the one-point vector amplitude $T_{\mu }^{V}$.
According to the standard definition this function contains one fermionic
propagator and the vertex operator $\Gamma _{i}=\gamma _{\mu }$. It is
defined, following our notation, as 
\begin{equation}
T_{\mu }^{V}=\int \frac{d^{4}k}{(2\pi )^{4}}Tr\left\{ \gamma _{\mu }\frac{1}{%
(\not{k}+\not{k}_{1})-m}\right\} ,
\end{equation}
and, after the trace is taken, we get 
\begin{equation}
T_{\mu }^{V}=4\left\{ \int \frac{d^{4}k}{(2\pi )^{4}}\frac{k_{\mu }}{%
(k+k_{1})^{2}-m^{2}}+k_{1\mu }\int \frac{d^{4}k}{(2\pi )^{4}}\frac{1}{%
(k+k_{1})^{2}-m^{2}}\right\} .
\end{equation}
From the above equation we see that we get two divergent integrals, one of
cubic and the other of quadratic degree of divergences. Following our
strategy we admit the presence of an implicit regulator, as discussed
before. In order to indicate its presence we use the subscript $\Lambda $ in
the integral and proceed to the necessary manipulations of the integrand in
such a way that the dependence on $k_{1}$ momentum is present only in finite
integrals in which case the connection limit is invocated. This goal can be
achieved by the use of the identity eq.(49) choosing $N$ not minor than
three for the cubic divergent integral above. Proceeding on this way we get: 
\begin{eqnarray}
\int_{\Lambda }\frac{d^{4}k}{(2\pi )^{4}}\frac{k_{\mu }}{(k+k_{1})^{2}-m^{2}}
&=&-k_{1\nu }\left\{ \int_{\Lambda }\frac{d^{4}k}{(2\pi )^{4}}\frac{2k_{\mu
}k_{\nu }}{(k^{2}-m^{2})^{2}}\right\}  \nonumber \\
&&+k_{1\nu }k_{1\alpha }k_{1\beta }\left\{ \int_{\Lambda }\frac{d^{4}k}{%
(2\pi )^{4}}\frac{4g_{\alpha \beta }k_{\mu }k_{\nu }}{(k^{2}-m^{2})^{3}}%
-\int_{\Lambda }\frac{d^{4}k}{(2\pi )^{4}}\frac{8k_{\alpha }k_{\beta }k_{\mu
}k_{\nu }}{(k^{2}-m^{2})^{4}}\right\}  \nonumber \\
&&-\left\{ \int \frac{d^{4}k}{(2\pi )^{4}}\frac{6k_{1}^{4}k_{1\nu }k_{\mu
}k_{\nu }}{(k^{2}-m^{2})^{4}}-\int \frac{d^{4}k}{(2\pi )^{4}}\frac{%
(k_{1}^{2}+2k_{1}\cdot k)^{4}k_{\mu }}{(k^{2}-m^{2})^{4}[(k+k_{1})^{2}-m^{2}]%
}\right\} .
\end{eqnarray}
Note the absence of the odd integrals in consequence of the even character
of our implicit regulator and that we have removed the subscript $\Lambda $
on the two last integrals in consequence of the connection limit
requirement. The finite integrals so obtained can be solved by standard
techniques without any restriction. The result is an exact cancellation
between them.

For the quadratic divergent integral in eq.(56) we apply the same recipe.
Choosing in the eq.(49) the value $N=2$ we get 
\begin{eqnarray}
\int_{\Lambda }\frac{d^{4}k}{(2\pi )^{4}}\frac{1}{(k+k_{1})^{2}-m^{2}}
&=&\int_{\Lambda }\frac{d^{4}k}{(2\pi )^{4}}\frac{1}{(k^{2}-m^{2})} 
\nonumber \\
&&+k_{1\mu }k_{1\nu }\left\{ \int_{\Lambda }\frac{d^{4}k}{(2\pi )^{4}}\frac{%
4k_{\mu }k_{\nu }}{(k^{2}-m^{2})^{3}}-\int_{\Lambda }\frac{d^{4}k}{(2\pi
)^{4}}\frac{g_{\mu \nu }}{(k^{2}-m^{2})^{2}}\right\}  \nonumber \\
&&+\left\{ \int \frac{d^{4}k}{(2\pi )^{4}}\frac{k_{1}^{4}}{(k^{2}-m^{2})^{3}}%
-\int \frac{d^{4}k}{(2\pi )^{4}}\frac{(k_{1}^{2}+2k_{1}\cdot k)^{3}}{%
(k^{2}-m^{2})^{3}[(k+k_{1})^{2}-m^{2}]}\right\} ,
\end{eqnarray}
where, again, the last two integrals cancel out after that the integration
have been performed. Then, collecting the results of eq.(57) and eq.(58)
without any modification we get for the $T_{\mu }^{V}$ one-point function 
\[
T_{\mu }^{V}=4\left\{ -k_{1\beta }\nabla _{\beta \mu }-\frac{k_{1\beta
}k_{1\alpha }k_{1\nu }}{3}\Box _{\alpha \beta \mu \nu }+\frac{%
k_{1}^{2}k_{1\nu }}{3}\triangle _{\nu \mu }+\frac{2}{3}k_{1\mu }k_{1\alpha
}k_{1\beta }\triangle _{\alpha \beta }\right\} , 
\]
where we have introduced a set of differences between divergent integrals
with the same degree of divergence, defined in eqs.(50)-(52). As announced
on the preceding comments we will analyze the results only after all
calculations have been performed. For while we only call the attention for
the very general character of the assumptions we have made to produce the
result eq.(59). No specific assumptions about a regulator are used and no
shift, limits, expansions and so on involving the divergent parts are
assumed. This is due to that, in fact, no calculations of divergent
integrals have been made.

Let us now treat the remaining one-point functions within the same
calculational scheme. Taking the scalar one, defined by 
\begin{equation}
T^{S}=\int \frac{d^{4}k}{(2\pi )^{4}}Tr\left\{ \hat{1}\frac{1}{(\not{k}+\not{%
k}_{1})-m}\right\}
\end{equation}
we get, after Dirac's trace, 
\begin{equation}
T^{S}=4m\int \frac{d^{4}k}{(2\pi )^{4}}\frac{1}{(k+k_{1})^{2}-m^{2}}.
\end{equation}
The quadratically divergent integral in the above expression has already
been discussed in the previous calculation. In consequence the one-point
function $T^{S}$ can be written as 
\begin{equation}
T^{S}=4m\left\{ I_{quad}(m^{2})+k_{1\beta }k_{1\alpha }\triangle _{\beta
\alpha }\right\} .
\end{equation}
In the above result we verify that the amplitude $T^{S}$ is expressed in
terms of two divergent objects, $I_{quad}(m^{2})$ and $\triangle _{\beta
\alpha }$, the latter again related to the arbitrary momentum label $k_{1}$.
For the moment we only comment that the two calculated amplitudes are 
\underline{potentially} ambiguous due to the presence of such terms in the
final expression.

It is a simple matter to check that the remaining one-point functions $%
T_{\mu }^{A}$ and $T^{P}$ vanish identically due to the trace's properties
of the Dirac's matrices involved.

\subsection{Two-Point Functions}

Let us calculate the two-point functions within the same scheme. We start by
considering the simplest one of such set of functions, namely, the
scalar-scalar, defined by 
\begin{equation}
T^{SS}=\int \frac{d^{4}k}{(2\pi )^{4}}Tr\left\{ \hat{1}\frac{1}{(\not{k}+{%
\not{k}}_{1})-m}\hat{1}\frac{1}{(\not{k}+{\not{k}}_{2})-m}\right\} ,
\end{equation}
where $k_{1}$ and $k_{2}$ stand for the arbitrary choices for the internal
momenta. Only the difference between them is a physical quantity, the
external momentum, consequently other combinations than the difference in
the result for the two-point functions will systematically locate
ambiguities, wherever the case may be. After taking Dirac traces and some
standard algebraic manipulations we get 
\begin{eqnarray}
T^{SS} &=&2\left\{ \int \frac{d^{4}k}{(2\pi )^{4}}\frac{1}{%
(k+k_{1})^{2}-m^{2}}+\int \frac{d^{4}k}{(2\pi )^{4}}\frac{1}{%
(k+k_{2})^{2}-m^{2}}\right.  \nonumber \\
&&\left. +[4m^{2}-(k_{1}-k_{2})^{2}]\int \frac{d^{4}k}{(2\pi )^{4}}\frac{1}{%
[(k+k_{1})^{2}-m^{2}][(k+k_{2})^{2}-m^{2}]}\right\} ,
\end{eqnarray}
where we have identified the quadratically divergent integrals which already
appeared in previous calculations. As for the second, logarithmically
divergent, we apply the appropriate manipulations to cast it into the form: 
\begin{eqnarray}
\int_{\Lambda }\frac{d^{4}k}{(2\pi )^{4}}\frac{1}{%
[(k+k_{1})^{2}-m^{2}][(k+k_{2})^{2}-m^{2}]} &=&\int_{\Lambda }\frac{d^{4}k}{%
(2\pi )^{4}}\frac{1}{(k^{2}-m^{2})^{2}}  \nonumber \\
&&-\int \frac{d^{4}k}{(2\pi )^{4}}\frac{(k_{1}^{2}+2k_{1}\cdot k)}{%
[(k^{2}-m^{2})^{2}][(k+k_{1})^{2}-m^{2}]}  \nonumber \\
&&-\int \frac{d^{4}k}{(2\pi )^{4}}\frac{(k_{2}^{2}+2k_{2}\cdot k)}{%
[(k^{2}-m^{2})^{2}][(k+k_{2})^{2}-m^{2}]}  \nonumber \\
&&+\int \frac{d^{4}k}{(2\pi )^{4}}\frac{(k_{1}^{2}+2k_{1}\cdot
k)(k_{2}^{2}+2k_{2}\cdot k)}{%
[(k^{2}-m^{2})^{2}][(k+k_{1})^{2}-m^{2}][(k+k_{2})^{2}-m^{2}]},
\end{eqnarray}
where we chosen, in the eq.(49), $N=1$ for the two propagators involved
which is very convenient, instead not unique, since it maintains the
symmetry in $k_{1}$ and $k_{2}$. The divergent content of this amplitude is
contained in the basic divergent object $I_{log}$. The remaining integrals
are finite and yield 
\begin{equation}
\int \frac{d^{4}k}{(2\pi )^{4}}\frac{(k_{1}^{2}+2k_{1}\cdot k)}{%
[(k^{2}-m^{2})^{2}][(k+k_{1})^{2}-m^{2}]}=\left( \frac{i}{(4\pi )^{2}}%
\right) [Z_{0}(m^{2},m^{2},k_{1}^{2};m^{2})],
\end{equation}
and 
\begin{eqnarray}
&&\!\!\!\!\!\!\!\!\!\int \frac{d^{4}k}{(2\pi )^{4}}\frac{(k_{1}^{2}+2k_{1}%
\cdot k)(k_{2}^{2}+2k_{2}\cdot k)}{%
[(k^{2}-m^{2})^{2}][(k+k_{1})^{2}-m^{2}][(k+k_{2})^{2}-m^{2}]}=  \nonumber \\
&&\!\!\!\!\!\!\left( \frac{i}{(4\pi )^{2}}\right) \left[
Z_{0}(m^{2},m^{2},k_{1}^{2};m^{2})+Z_{0}(m^{2},m^{2},k_{2}^{2};m^{2})-Z_{0}(m^{2},m^{2},(k_{1}-k_{2})^{2};m^{2})%
\right] ,
\end{eqnarray}
where we leave the integration in the last of Feynman parameters through the
introduction of the structure functions for one loop integrals defined as 
\cite{ref19} 
\begin{equation}
Z_{k}(\lambda _{1}^{2},\lambda _{2}^{2},q^{2};\lambda
^{2})=\int_{0}^{1}dzz^{k}ln\left( \frac{q^{2}z(1-z)+(\lambda
_{1}^{2}-\lambda _{2}^{2})z-\lambda _{1}^{2}}{(-\lambda ^{2})}\right) ,
\end{equation}
which have been proven very useful in the systematization of this type of
calculations. Explicit expressions for $Z_{k}$ functions can be easily
obtained but are not relevant on this investigations. Collecting the
results, the logarithmically divergent integral can be written as 
\begin{equation}
\int \frac{d^{4}k}{(2\pi )^{4}}\frac{1}{%
[(k+k_{1})^{2}-m^{2}][(k+k_{2})^{2}-m^{2}]}=I_{log}(m^{2})-\left( \frac{i}{%
(4\pi )^{2}}\right) Z_{0}(m^{2},m^{2},(k_{1}-k_{2})^{2};m^{2})
\end{equation}
and, consequently, the $T^{SS}$ amplitude is given by 
\begin{eqnarray}
T^{SS} &=&4\left\{ I_{quad}(m^{2})+\frac{4m^{2}-(k_{1}-k_{2})^{2}}{2}%
[I_{log}(m^{2})]\right.  \nonumber \\
&&\left. -\frac{4m^{2}-(k_{1}-k_{2})^{2}}{2}\left( \frac{i}{(4\pi )^{2}}%
\right) Z_{0}(m^{2},m^{2},(k_{1}-k_{2})^{2};m^{2})\right\}  \nonumber \\
&&+\left[ (k_{1}-k_{2})_{\alpha }(k_{1}-k_{2})_{\beta }\right] \triangle
_{\alpha \beta }  \nonumber \\
&&+\left[ (k_{1}+k_{2})_{\alpha }(k_{1}+k_{2})_{\beta }\right] \triangle
_{\alpha \beta },
\end{eqnarray}
where the last term is, in principle, ambiguous. The finite contributions
and that proportional to the basic divergent integral $I_{log}$ are
unambiguous since the combination $k_{2}-k_{1}$ is just equal to the
external momentum.

With the same ingredients, rigorously, we can calculate the
pseudoscalar-pseudoscalar two-point function: 
\begin{eqnarray}
T^{PP} &=&\int \frac{d^{4}k}{(2\pi )^{4}}Tr\left\{ \gamma _{5}\frac{1}{(\not{%
k}+{\not{k}}_{1})-m}\gamma _{5}\frac{1}{(\not{k}+{\not{k}}_{2})-m}\right\} 
\nonumber \\
&=&4\left\{ -I_{quad}(m^{2})+\frac{(k_{1}-k_{2})^{2}}{2}[I_{log}(m^{2})]%
\right.  \nonumber \\
&&\;\;\;\;\left. -\frac{(k_{1}-k_{2})^{2}}{2}\left( \frac{i}{(4\pi )^{2}}%
\right) Z_{0}(m^{2},m^{2},(k_{1}-k_{2})^{2};m^{2})\right\}  \nonumber \\
&&\;\;\;\;-\left[ (k_{1}-k_{2})_{\alpha }(k_{1}-k_{2})_{\beta }\right]
\triangle _{\alpha \beta }  \nonumber \\
&&\;\;\;\;-\left[ (k_{1}+k_{2})_{\alpha }(k_{1}+k_{2})_{\beta }\right]
\triangle _{\alpha \beta },
\end{eqnarray}
which also presents a potentially ambiguous term. Now, for the evaluation of
the $T_{\mu }^{AP}$ two-point function only the result in eq.(69) is needed.
After the traces calculations: 
\begin{eqnarray}
T_{\mu }^{AP} &=&\int \frac{d^{4}k}{(2\pi )^{4}}Tr\left\{ i\gamma _{\mu
}\gamma _{5}\frac{1}{(\not{k}+{\not{k}}_{1})-m}\gamma _{5}\frac{1}{(\not{k}+{%
\not{k}}_{2})-m}\right\}  \nonumber \\
&=&-4mi(k_{1}-k_{2})_{\mu }\left\{ [I_{log}(m^{2})]-\left( \frac{i}{(4\pi
)^{2}}\right) Z_{0}(m^{2},m^{2},(k_{1}-k_{2})^{2};m^{2})\right\} ,
\end{eqnarray}
which is unambiguous. Following the calculations we consider the
vector-scalar two-point function, defined as 
\begin{equation}
T_{\mu }^{VS}=\int \frac{d^{4}k}{(2\pi )^{4}}Tr\left\{ \gamma _{\mu }\frac{1%
}{(\not{k}+{\not{k}}_{1})-m}\hat{1}\frac{1}{(\not{k}+{\not{k}}_{2})-m}%
\right\} ,
\end{equation}
which, solving the trace operation, can be written as 
\begin{eqnarray}
T_{\mu }^{VS} &=&4m\left\{ \int \frac{d^{4}k}{(2\pi )^{4}}\frac{2k_{\mu }}{%
[(k+k_{1})^{2}-m^{2}][(k+k_{2})^{2}-m^{2}]}\right.  \nonumber \\
&&\left. +(k_{1}+k_{2})_{\mu }\int \frac{d^{4}k}{(2\pi )^{4}}\frac{1}{%
[(k+k_{1})^{2}-m^{2}][(k+k_{2})^{2}-m^{2}]}\right\} ,
\end{eqnarray}
where a new type of divergent integral, linearly divergent, have appeared.

Following our recipe, we first use the identity eq.(49) with $N=1$ for both
terms in the integral to obtain 
\begin{eqnarray}
\int_{\Lambda }\frac{d^{4}k}{(2\pi )^{4}}\frac{k_{\mu }}{%
[(k+k_{1})^{2}-m^{2}][(k+k_{2})^{2}-m^{2}]} &=&-\frac{(k_{1}+k_{2})_{\alpha }%
}{2}\int_{\Lambda }\frac{d^{4}k}{(2\pi )^{4}}\frac{4k_{\alpha }k_{\mu }}{%
(k^{2}-m^{2})^{3}}  \nonumber \\
&&+\int \frac{d^{4}k}{(2\pi )^{4}}\frac{(k_{1}^{2}+2k_{1}\cdot k)^{2}k_{\mu }%
}{[(k^{2}-m^{2})^{3}][(k+k_{1})^{2}-m^{2}]}  \nonumber \\
&&+\int \frac{d^{4}k}{(2\pi )^{4}}\frac{(k_{2}^{2}+2k_{2}\cdot k)^{2}k_{\mu }%
}{[(k^{2}-m^{2})^{3}][(k+k_{2})^{2}-m^{2}]}  \nonumber \\
\!\!\!\!\!\!\!\!\!\!\!\!\!\!\!\!\!\!\!\!\!\!\! &&+\int \frac{d^{4}k}{(2\pi
)^{4}}\frac{(k_{1}^{2}+2k_{1}\cdot k)(k_{2}^{2}+2k_{2}\cdot k)k_{\mu }}{%
[(k^{2}-m^{2})^{2}][(k+k_{1})^{2}-m^{2}][(k+k_{2})^{2}-m^{2}]}.
\end{eqnarray}
Only an odd integral have been removed on the right hand side of the above
equation, in addition to the subscript $\Lambda $ on the last three (finite)
integrals, which, after the integration, produces the results; 
\begin{equation}
\bullet \int \frac{d^{4}k}{(2\pi )^{4}}\frac{(k_{1}^{2}+2k_{1}\cdot
k)^{2}k_{\mu }}{[(k^{2}-m^{2})^{3}][(k+k_{1})^{2}-m^{2}]}=\left( \frac{i}{%
(4\pi )^{2}}\right) [k_{1\mu }][Z_{1}(m^{2},m^{2},k_{1}^{2};m^{2})]
\end{equation}

\begin{eqnarray}
&&\!\!\!\!\!\!\!\!\!\!\!\!\bullet \int \frac{d^{4}k}{(2\pi )^{4}}\frac{%
(k_{1}^{2}+2k_{1}\cdot k)(k_{2}^{2}+2k_{2}\cdot k)k_{\mu }}{%
[(k^{2}-m^{2})^{2}][(k+k_{1})^{2}-m^{2}][(k+k_{2})^{2}-m^{2}]}=  \nonumber \\
&&\;\;\;\;\;\;\;\;\;\;=\left( \frac{i}{(4\pi )^{2}}\right) \left[ -k_{1\mu
}Z_{1}(m^{2},m^{2},k_{1}^{2};m^{2})-k_{2\mu
}Z_{1}(m^{2},m^{2},k_{2}^{2};m^{2})\right.  \nonumber \\
&&\;\;\;\;\;\;\;\;\;\;\;\;\;\;\left. +(k_{1}+k_{2})_{\mu
}Z_{1}(m^{2},m^{2},(k_{1}-k_{2})^{2};m^{2})\right] .
\end{eqnarray}
If we consider also a property of the $Z_{k}$ functions 
\begin{equation}
Z_{1}(m^{2},m^{2},q^{2};m^{2})=\frac{Z_{0}(m^{2},m^{2},q^{2};m^{2})}{2},
\end{equation}
we can write the result for the linearly divergent integral as: 
\begin{eqnarray}
\int_{\Lambda }\frac{d^{4}k}{(2\pi )^{4}}\frac{k_{\mu }}{%
[(k+k_{1})^{2}-m^{2}][(k+k_{2})^{2}-m^{2}]} &=&-\frac{(k_{1}+k_{2})_{\alpha }%
}{2}\int_{\Lambda }\frac{d^{4}k}{(2\pi )^{4}}\frac{4k_{\mu }k_{\alpha }}{%
(k^{2}-m^{2})^{3}}  \nonumber \\
&&\!\!\!\!\!\!\!\!\!\!\!\!\!\!\!\!\!\!\!\!\!\!\!\!\!\!\!\!\!\!\!\!\!\!\!\!\!%
\!\!\!\!\!\!\!\!\!\!\!\!\!\!\!\!\!\!+\left( \frac{i}{(4\pi )^{2}}\right) 
\frac{(k_{1}+k_{2})_{\mu }}{2}[Z_{0}(m^{2},m^{2},(k_{1}-k_{2})^{2};m^{2})].
\end{eqnarray}
Then, after the substitution of the eqs.(69) and (79) results, we get for
the $T_{\mu }^{VS}$ amplitude 
\begin{equation}
T_{\mu }^{VS}=(-)4m(k_{1}+k_{2})_{\beta }[\triangle _{\beta \mu }].
\end{equation}

Let us now consider those two-point functions with two Lorentz indexes.
First the one which play a crucial role in our investigation, as become
clear in what follows, namely the axial-vector-vector, defined by 
\begin{equation}
T_{\mu \nu }^{AV}=\int \frac{d^{4}k}{(2\pi )^{4}}Tr\left\{ i\gamma _{\mu
}\gamma _{5}\frac{1}{(\not{k}+{\not{k}}_{1})-m}\gamma _{\nu }\frac{1}{(\not{k%
}+{\not{k}}_{2})-m}\right\} .
\end{equation}
Solving the traces involved; 
\begin{eqnarray}
T_{\mu \nu }^{AV} &=&-4\varepsilon _{\mu \nu \alpha \beta }\left\{ k_{2\beta
}\int \frac{d^{4}k}{(2\pi )^{4}}\frac{k_{\alpha }}{%
[(k+k_{1})^{2}-m^{2}][(k+k_{2})^{2}-m^{2}]}\right.  \nonumber \\
&&+k_{1\alpha }\int \frac{d^{4}k}{(2\pi )^{4}}\frac{k_{\beta }}{%
[(k+k_{1})^{2}-m^{2}][(k+k_{2})^{2}-m^{2}]}  \nonumber \\
&&\left. +k_{1\alpha }k_{2\beta }\int \frac{d^{4}k}{(2\pi )^{4}}\frac{1}{%
[(k+k_{1})^{2}-m^{2}][(k+k_{2})^{2}-m^{2}]}\right\} .
\end{eqnarray}
To complete the calculations only the results eq.(69) and Eq.(79), the same
ones used in the $T_{\mu }^{VS}$ calculations, are needed. We obtain then 
\begin{equation}
T_{\mu \nu }^{AV}=2\varepsilon _{\mu \nu \alpha \beta }\left[
(k_{2}-k_{1})_{\beta }(k_{1}+k_{2})_{\xi }\triangle _{\xi \alpha }\right] .
\end{equation}
Note that this amplitude constitutes the first pseudo one considered at this
point. For while we call attention for the fact that no special procedures
have been adopted.

Remain to evaluate only two amplitudes, the more complex ones. First we take
the $T_{\mu \nu }^{VV}$ amplitude defined by 
\begin{equation}
T_{\mu \nu }^{VV}=\int \frac{d^{4}k}{(2\pi )^{4}}Tr\left\{ \gamma _{\mu }%
\frac{1}{(\not{k}+{\not{k}}_{1})-m}\gamma _{\nu }\frac{1}{(\not{k}+{\not{k}}%
_{2})-m}\right\} .
\end{equation}
The traces evaluations lead us to write, 
\begin{equation}
T_{\mu \nu }^{VV}=T_{\mu \nu }+g_{\mu \nu }[T^{PP}],
\end{equation}
where the term proportional to $g_{\mu \nu }$, coming from the trace, is an
identical mathematical structure to the $T^{PP}$ two-point function, in such
a way the same results will be adopted, and the tensor $T_{\mu \nu }$ is
defined by 
\begin{equation}
T_{\mu \nu }=4\int \frac{d^{4}k}{(2\pi )^{4}}\frac{[(k+k_{1})_{\mu
}(k+k_{2})_{\nu }+(k+k_{1})_{\nu }(k+k_{2})_{\mu }]}{%
[(k+k_{1})^{2}-m^{2}][(k+k_{2})^{2}-m^{2}]}.
\end{equation}
To evaluate $T_{\mu \nu }$ we first need to consider the quadratically
divergent integral present in the above expression. Following strictly the
same procedure adopted on the evaluation of the logarithmically and linearly
divergent cases, choosing adequate values for $N$ in the identity eq.(49),
after a long and tedious calculations, we will arrive to the result 
\begin{eqnarray}
&&\int_{\Lambda }\frac{d^{4}k}{(2\pi )^{4}}\frac{k_{\mu }k_{\nu }}{%
[(k+k_{1})^{2}-m^{2}][(k+k_{2})^{2}-m^{2}]}=  \nonumber \\
&&\;\;\;\;\;\;\;\;\;\;\;\;\;\;\;\;\;\;\;\;\;\;\;\left\{ \int_{\Lambda }\frac{%
d^{4}k}{(2\pi )^{4}}\frac{k_{\mu }k_{\nu }}{[(k^{2}-m^{2})^{2}}\right\} 
\nonumber \\
&&\;\;\;\;\;\;\;\;\;\;\;\;\;\;\;\;\;\;\;\;\;\;\;+\left\{
-(k_{1}^{2}+k_{2}^{2})\int_{\Lambda }\frac{d^{4}k}{(2\pi )^{4}}\frac{k_{\mu
}k_{\nu }}{[(k^{2}-m^{2})^{3}}+(k_{1\xi }k_{1\beta }+k_{2\xi }k_{2\beta
}+k_{1\beta }k_{2\xi })\int_{\Lambda }\frac{d^{4}k}{(2\pi )^{4}}\frac{%
4k_{\xi }k_{\beta }k_{\mu }k_{\nu }}{[(k^{2}-m^{2})^{4}}\right\}  \nonumber
\\
&&\;\;\;\;\;\;\;\;\;\;\;\;\;\;\;\;\;\;\;\;\;\;\;+\left. \left( \frac{i}{%
(4\pi )^{2}}\right) \right\{ \left[ (k_{1}-k_{2})_{\mu }(k_{1}-k_{2})_{\nu
}-(k_{1}-k_{2})^{2}g_{\mu \nu }\right] \times  \nonumber \\
&&\;\;\;\;\;\;\;\;\;\;\;\;\;\;\;\;\;\;\;\;\;\;\;\;\;\;\;\;\;\;\;\;\;\;\;\;\;%
\;\;\;\;\;\;\;\;\;\times \left[ -Z_{2}(m^{2},m^{2},(k_{1}-k_{2})^{2};m^{2})+%
\frac{Z_{0}(m^{2},m^{2},(k_{1}-k_{2})^{2};m^{2})}{4}\right]  \nonumber \\
&&\;\;\;\;\;\;\;\;\;\;\;\;\;\;\;\;\;\;\;\;\;\;\;\;\;\;\;\;\;\;\;\;\;\;\;\;\;%
\;\;\;\;\;\;\;\;\;-\left. \frac{(k_{1}+k_{2})_{\mu }(k_{1}+k_{2})_{\nu }}{4}%
Z_{0}(m^{2},m^{2},(k_{1}-k_{2})^{2};m^{2})\right\} .
\end{eqnarray}
With this result in addition to eq.(69), eq.(79), eq.(85), eq.(71) and the
property of the $Z_{k}$ functions 
\begin{equation}
Z_{2}(m^{2},m^{2},q^{2};m^{2})=-\frac{1}{18}-\frac{m^{2}}{3q^{2}}%
Z_{0}(m^{2},m^{2},q^{2};m^{2})+\frac{Z_{0}(m^{2},m^{2},q^{2};m^{2})}{3},
\end{equation}
a very useful expression for $T_{\mu \nu }$ can be obtained 
\begin{eqnarray}
T_{\mu \nu } &=&\frac{4}{3}[(k_{1}-k_{2})^{2}g_{\mu \nu }-(k_{1}-k_{2})_{\mu
}(k_{1}-k_{2})_{\nu }]\times  \nonumber \\
&&\times \left\{ I_{log}(m^{2})-\left( \frac{i}{(4\pi )^{2}}\right) \left[ 
\frac{1}{3}+\left( \frac{2m^{2}+(k_{1}-k_{2})^{2}}{(k_{1}-k_{2})^{2}}\right)
Z_{0}(m^{2},m^{2},(k_{1}-k_{2})^{2};m^{2})\right] \right\}  \nonumber \\
&&-[T^{PP}]g_{\mu \nu }+A_{\mu \nu },
\end{eqnarray}
where 
\begin{eqnarray}
A_{\mu \nu } &=&4[\nabla _{\mu \nu }]+(k_{1}-k_{2})_{\alpha
}(k_{1}-k_{2})_{\beta }\left[ \frac{1}{3}\Box _{\alpha \beta \mu \nu }+\frac{%
1}{3}\triangle _{\mu \beta }g_{\alpha \nu }+g_{\alpha \mu }\triangle _{\beta
\nu }-g_{\mu \nu }\triangle _{\alpha \beta }-\frac{2}{3}g_{\alpha \beta
}\triangle _{\mu \nu }\right]  \nonumber \\
&&+\left[ (k_{1}-k_{2})_{\alpha }(k_{1}+k_{2})_{\beta
}-(k_{1}+k_{2})_{\alpha }(k_{1}-k_{2})_{\beta }\right] \left[ \frac{1}{3}%
\Box _{\alpha \beta \mu \nu }+\frac{1}{3}\triangle _{\mu \beta }g_{\nu
\alpha }+\frac{1}{3}\triangle _{\beta \nu }g_{\alpha \mu }\right]  \nonumber
\\
&&+(k_{1}+k_{2})_{\alpha }(k_{1}+k_{2})_{\beta }\left[ \Box _{\alpha \beta
\mu \nu }-\triangle _{\nu \alpha }g_{\mu \beta }-\triangle _{\beta \nu
}g_{\alpha \mu }-3\triangle _{\alpha \beta }g_{\mu \nu }\right] .
\end{eqnarray}
Then we are at the position to write the expression for the $T_{\mu \nu
}^{VV}$ amplitude 
\begin{eqnarray}
T_{\mu \nu }^{VV} &=&\frac{4}{3}[(k_{1}-k_{2})^{2}g_{\mu \nu
}-(k_{1}-k_{2})_{\mu }(k_{1}-k_{2})_{\nu }]\times  \nonumber \\
&&\times \left\{ \lbrack I_{log}(m^{2})]-\left( \frac{i}{(4\pi )^{2}}\right) %
\left[ \frac{1}{3}+\frac{(2m^{2}+(k_{1}-k_{2})^{2})}{(k_{1}-k_{2})^{2}}%
Z_{0}(m^{2},m^{2},(k_{1}-k_{2})^{2};m^{2})\right] \right\} +A_{\mu \nu }.
\end{eqnarray}

Finally we consider the $AA$ two-point function, 
\begin{equation}
T_{\mu \nu }^{AA}=\int \frac{d^{4}k}{(2\pi )^{4}}Tr\left\{ i\gamma _{\mu
}\gamma _{5}\frac{1}{[\not{k}+\not{k}_{1}-m]}i\gamma _{\nu }\gamma _{5}\frac{%
1}{[\not{k}+\not{k}_{2}-m]}\right\} ,
\end{equation}
which exhibits the structure 
\begin{equation}
T_{\mu \nu }^{AA}=g_{\mu \nu }[T^{SS}]-T_{\mu \nu }.
\end{equation}
After the traces evaluation is performed. No additional calculations are
needed if we identify the relation between two-point functions 
\begin{equation}
\left[ T^{SS}+T^{PP}\right] =\frac{2mi(k_{1}-k_{2})_{\mu }}{(k_{1}-k_{2})^{2}%
}T_{\mu }^{AP},
\end{equation}
collecting the results eq.(89), eq.(94), and substituting in eq.(93), the
expression for $T_{\mu \nu }^{AA}$ amplitude so obtained is: 
\begin{eqnarray}
T_{\mu \nu }^{AA} &=&-\frac{4}{3}[(k_{1}-k_{2})^{2}g_{\mu \nu
}-(k_{1}-k_{2})_{\mu }(k_{1}-k_{2})_{\nu }]\times  \nonumber \\
&&\times \left\{ \lbrack I_{log}(m^{2})]-\left( \frac{i}{(4\pi )^{2}}\right) %
\left[ \frac{1}{3}+\frac{(2m^{2}+(k_{1}-k_{2})^{2})}{(k_{1}-k_{2})^{2}}%
Z_{0}(m^{2},m^{2},(k_{1}-k_{2})^{2};m^{2})\right] \right\}  \nonumber \\
&&+g_{\mu \nu }8m^{2}\left\{ [I_{log}(m^{2})]-\left( \frac{i}{(4\pi )^{2}}%
\right) Z_{0}(m^{2},m^{2},(k_{1}-k_{2})^{2};m^{2})\right\} -A_{\mu \nu }.
\end{eqnarray}
This calculation completes the evaluation of all non-vanishing two-point
functions. It is a simple matter to show that the remaining ones $T_{\mu
}^{VP}$, $T^{PS}$ and $T_{\mu }^{AS}$ vanishes identically due to the
presence of the $\gamma _{5}$ matrix.

\subsection{Three Point Functions}

We now come to the three-point functions. In order not to overload the text
we limit ourselves in writing, in the explicit way, only the terms
indicating possibility of ambiguities. This is sufficient for the analysis
we want to do, closely related to that performed in ref.\cite{ref1}. To
illustrate what we mean, consider a relatively simple case of three-point
functions: 
\begin{equation}
T_{\lambda }^{VSS}=\int \frac{d^{4}k}{(2\pi )^{4}}Tr\left\{ \gamma _{\lambda
}\frac{1}{[\not{k}+\not{k}_{3}-m]}\hat{1}\frac{1}{[\not{k}+\not{k}_{2}-m]}%
\hat{1}\frac{1}{[\not{k}+\not{k}_{1}-m]}\right\}
\end{equation}
After taking the trace and performing some standard algebraic reorganization
of terms we obtain the expression 
\begin{eqnarray}
T_{\lambda }^{VSS} &=&2\left\{ \int \frac{d^{4}k}{(2\pi )^{4}}\frac{%
2k_{\lambda }}{[(k+k_{2})^{2}-m^{2}][(k+k_{3})^{2}-m^{2}]}\right.  \nonumber
\\
&&+(k_{2}+k_{3})_{\lambda }\int \frac{d^{4}k}{(2\pi )^{4}}\frac{1}{%
[(k+k_{2})^{2}-m^{2}][(k+k_{3})^{2}-m^{2}]}  \nonumber \\
&&+(k_{3}-k_{1})_{\lambda }\int \frac{d^{4}k}{(2\pi )^{4}}\frac{1}{%
[(k+k_{1})^{2}-m^{2}][(k+k_{3})^{2}-m^{2}]}  \nonumber \\
&&+(k_{2}-k_{1})_{\lambda }\int \frac{d^{4}k}{(2\pi )^{4}}\frac{1}{%
[(k+k_{1})^{2}-m^{2}][(k+k_{2})^{2}-m^{2}]}  \nonumber \\
&&+\left[ 8m^{2}-(k_{1}-k_{2})^{2}-(k_{1}-k_{3})^{2}+(k_{2}-k_{3})^{2}\right]
\times  \nonumber \\
&&\;\;\;\;\;\times \int \frac{d^{4}k}{(2\pi )^{4}}\frac{k_{\lambda }}{%
[k^{2}-m^{2}][(k+k_{2}-k_{1})^{2}-m^{2}][(k+k_{3}-k_{1})^{2}-m^{2}]} 
\nonumber \\
&&+\left[ [4m^{2}-(k_{1}-k_{2})^{2}](k_{3}-k_{1})_{\lambda
}+[4m^{2}-(k_{1}-k_{3})^{2}](k_{2}-k_{1})_{\lambda }\right] \times  \nonumber
\\
&&\;\;\;\;\;\left. \times \int \frac{d^{4}k}{(2\pi )^{4}}\frac{1}{%
[k^{2}-m^{2}][(k+k_{2}-k_{1})^{2}-m^{2}][(k+k_{3}-k_{1})^{2}-m^{2}]}\right\}
.
\end{eqnarray}
In the above expression we note that only the first two terms are
potentially ambiguous. The remaining ones are either logarithmically
divergent or finite, all with unambiguous coefficients. Of course it is
possible to solve all integrals and obtain an analytical expression for $%
T_{\lambda }^{VSS}$ \cite{ref19}. But, for our immediate purposes, it is
enough to write it in the form 
\begin{equation}
T_{\lambda }^{VSS}=-2(k_{2}+k_{3})_{\xi }[\triangle _{\lambda \xi }]+NAT,
\end{equation}
where NAT stands for non-ambiguous terms. Having this in mind we simply list
the other three-point functions in the same manner; 
\begin{equation}
T_{\lambda }^{VPP}=2(k_{2}+k_{3})_{\xi }[\triangle _{\lambda \xi }]+NAT,
\end{equation}
with only one Lorentz index we have also 
\begin{equation}
T_{\mu }^{SAP}=2i(k_{1}+k_{3})_{\xi }[\triangle _{\mu \xi }]+NAT.
\end{equation}
On the other hand, with three Lorentz indexes; 
\begin{equation}
T_{\lambda \mu \nu }^{AAA}=-2\varepsilon _{\lambda \mu \nu \xi
}(k_{1}+k_{2})_{\sigma }(\triangle _{\xi \sigma })+NAT,
\end{equation}

\begin{equation}
T^{AVV}_{\lambda\mu\nu}=2\varepsilon_{\lambda\mu\nu\xi}(k_1+k_2)_\sigma
(\triangle_{\xi\sigma})+NAT
\end{equation}
and

\begin{eqnarray}
T_{\lambda \mu \nu }^{VVV} &=&\left\{ (k_{1}+k_{2})_{\xi }\left[ -\frac{2}{3}%
\Box _{\xi \mu \nu \lambda }-\frac{2}{3}g_{\xi \nu }\triangle _{\mu \lambda
}-\frac{2}{3}g_{\xi \mu }\triangle _{\nu \lambda }\right. -\frac{2}{3}g_{\xi
\lambda }\triangle _{\mu \nu }+2g_{\lambda \mu }\triangle _{\nu \xi
}+2g_{\xi \nu }\triangle _{\lambda \mu }\right]  \nonumber \\
&&+(k_{1}+k_{3})_{\xi }\left[ -\frac{2}{3}\Box _{\xi \mu \nu \lambda }-\frac{%
2}{3}g_{\xi \nu }\triangle _{\mu \lambda }-\frac{2}{3}g_{\xi \mu }\triangle
_{\nu \lambda }-\frac{2}{3}g_{\xi \lambda }\triangle _{\mu \nu }+2g_{\lambda
\nu }\triangle _{\mu \xi }+2g_{\xi \mu }\triangle _{\lambda \nu }\right] 
\nonumber \\
&&\left. +(k_{2}+k_{3})_{\xi }\left[ -\frac{2}{3}\Box _{\xi \mu \nu \lambda
}-\frac{2}{3}g_{\xi \nu }\triangle _{\mu \lambda }-\frac{2}{3}g_{\xi \mu
}\triangle _{\nu \lambda }-\frac{2}{3}g_{\xi \lambda }\triangle _{\mu \nu
}+2g_{\mu \nu }\triangle _{\lambda \xi }+2g_{\xi \lambda }\triangle _{\mu
\nu }\right] +NAT\right\} .
\end{eqnarray}
In order to give a complete account of the three-point functions we give the
amplitude $T_{\lambda \mu \nu }^{VAA}$, which possesses the same potentially
ambiguous term (AT) as $T_{\lambda \mu \nu }^{VVV}$, i.e. 
\begin{equation}
\left. T_{\lambda \mu \nu }^{VAA}\right| _{AT}=-\left. T_{\lambda \mu \nu
}^{VVV}\right| _{AT}.
\end{equation}
The remaining three-point functions are all unambiguous, although divergent,
therefore are left out of the discussions.

Let us now establish contact between our results and those obtained by
Gertsein and Jackiw showing that it is a simple matter to go from our
formalism over to theirs.

\section{Ambiguities in Gertsein and Jackiw Approach}

Let us now understand how the results obtained in the previous section
include those of ref.\cite{ref1} as a special case. We show that starting
from our expressions we can obtain the results in tables I and II for the
ambiguities in the two and three point functions as defined by those authors
in section III of ref.\cite{ref1}. We start with our results for the
amplitude $T_{\mu }^{VS}$ given by eq.(80). In the adopted notation the
ambiguity has been introduced via the arbitrary momenta $k_{1}$ and $k_{2}$,
by calculating the amplitudes with internal momenta in the loop: $k+k_{1}$
and $k+k_{2}$ in the two propagators in question, as defined in eq.(6). In
ref.\cite{ref1} the arbitrariness is represented by the momentum $s$, such
that the propagators have momenta $k+s$ and $k+s+p$ which establish the
equivalent relation 
\[
\left\{ 
\begin{array}{c}
k_{1}=s \\ 
k_{2}=s+p.
\end{array}
\right. 
\]
In this way, the ambiguous part of the $T_{\mu }^{VS}$ amplitude can be cast
into the form 
\begin{equation}
T_{\mu }^{VS}=-4m(2s)_{\beta }\left\{ \int_{\Lambda }\frac{d^{4}k}{(2\pi
)^{4}}\frac{4k_{\mu }k_{\beta }}{(k^{2}-m^{2})^{3}}-\int_{\Lambda }\frac{%
d^{4}k}{(2\pi )^{4}}\frac{g_{\mu \beta }}{(k^{2}-m^{2})^{2}}\right\} ,
\end{equation}
where we have used the definition in eq.(51) for the object $\triangle
_{\beta \mu }$. Taking now the limit in a symmetric way in the first
integral, i.e., using the relation 
\begin{equation}
\lim_{k\rightarrow \infty }k_{\mu }k_{\beta }=\frac{1}{4}k^{2}g_{\mu \beta }
\end{equation}
and taking $k^{2}\rightarrow k^{2}-m^{2}+m^{2}$ in this integral we get 
\begin{equation}
T_{\mu }^{VS}=-4m(2s)_{\beta }\left\{ \int \frac{d^{4}k}{(2\pi )^{4}}\frac{%
m^{2}g_{\mu \beta }}{(k^{2}-m^{2})^{3}}\right\} .
\end{equation}
Using the result 
\begin{equation}
\int \frac{d^{4}k}{(2\pi )^{4}}\frac{1}{(k^{2}-m^{2})^{3}}=\frac{i}{2(4\pi
)^{2}(-m^{2})},
\end{equation}
we finally get 
\begin{equation}
T_{\mu }^{VS}=\frac{4i\pi ^{2}}{(2\pi )^{4}}ms_{\mu },
\end{equation}
as in ref.\cite{ref1}. In an analogous way we get for the ambiguous
contribution of $T^{PP}$ and $T^{SS}$ 
\begin{equation}
T^{PP}|_{AT}=T^{SS}|_{AT}=\frac{-i\pi ^{2}}{(2\pi )^{4}}s\cdot (s+p),
\end{equation}
and for $T_{\mu \nu }^{AV}$ 
\begin{equation}
T_{\mu \nu }^{AV}=\frac{-2i\pi ^{2}\varepsilon _{\mu \alpha \nu \beta
}s_{\alpha }p_{\beta }}{(2\pi )^{4}}.
\end{equation}
The functions $T_{\mu \nu }^{VV}$ and $T_{\mu \nu }^{AA}$ have the ambiguous
part given by $A_{\mu \nu }$ in eq.(90). An evaluation for $\Box _{\alpha
\beta \mu \nu }$, defined in eq.(50), is in order. It is necessary to use
the relation 
\begin{equation}
\lim_{k\rightarrow \infty }\frac{k_{\alpha }k_{\beta }k_{\mu }k_{\nu }}{k^{4}%
}=\frac{1}{24}\left( g_{\alpha \beta }g_{\mu \nu }+g_{\alpha \mu }g_{\beta
\nu }+g_{\alpha \nu }g_{\beta \mu }\right)
\end{equation}
and also the following results 
\begin{equation}
\int \frac{d^{4}k}{(2\pi )^{4}}\frac{k_{\mu }k_{\nu }}{(k^{2}-m^{2})^{4}}=%
\frac{i}{(4\pi )^{2}}\frac{1/2g_{\mu \nu }}{3!(-m^{2})}
\end{equation}
and 
\begin{equation}
\int \frac{d^{4}k}{(2\pi )^{4}}\frac{k^{2}}{(k^{2}-m^{2})^{4}}=\frac{i}{%
(4\pi )^{2}}\frac{1}{3(-m^{2})},
\end{equation}
in order to obtain (after some algebra) 
\begin{equation}
T_{\mu \nu }^{VV}|_{AT}=T_{\mu \nu }^{AA}|_{AT}=\frac{2i\pi ^{2}}{3(2\pi
)^{4}}\left\{ s_{\mu }(s+p)_{\nu }+g_{\mu \nu }s\cdot (s+p)+s_{\nu
}(s+p)_{\mu }\right\} .
\end{equation}

In the case of the three-point functions, in order to obtain the results of
ref.\cite{ref1} it is necessary to take the SU(3) indices into account in
the calculation of the amplitudes, since they play an important role in the
construction of currents. The Gell-Mann matrices introduce symmetry factors
such as relative global signs between the direct and crossed channel, which
are not present in our non-abelian model. However this does not introduce
any inconsistency in the two formulations, as we will see.

The arbitrariness in the label for the internal momenta of the loop is
introduced by taking 
\begin{equation}
s=ap+bp^{\prime },
\end{equation}
where $a$ and $b$ are real numbers. The ambiguity is then found by means of
calculating the difference between two representations for the three-point
functions: with and without the arbitrary label $s$ defined above. In this
sense we see that it can be interpreted as a surface term, which would
disappear if shifts in the integration variable were allowed.

Following this prescription, the ambiguous term of $T_{\mu }^{SAP}$
previously found in the form 
\begin{equation}
T_{\mu }^{SAP}|_{AT}=2i(k_{1}+k_{3})_{\xi }\triangle _{\mu \xi }
\end{equation}
for the direct channel (without SU(3) factors) may be written as 
\begin{equation}
T_{\mu }^{SAP}|_{AT}=\frac{2\pi ^{2}}{(2\pi )^{4}}(ap+bp^{\prime })_{\mu }
\end{equation}
which, after the inclusion of the crossed channel and the pertinent factors
yield 
\begin{equation}
T_{\mu }^{S\rightarrow AP}|_{AT}=\frac{2\pi ^{2}}{(2\pi )^{4}}%
(a-b)(p-p^{\prime })_{\mu }.
\end{equation}
In a completely analogous way we get 
\begin{equation}
T_{\mu }^{V\rightarrow SS}|_{AT}=T_{\mu }^{V\rightarrow PP}|_{AT}=\frac{%
-i2\pi ^{2}}{(2\pi )^{4}}(a-b)(p-p^{\prime })_{\mu },
\end{equation}
\begin{equation}
T_{\mu \alpha \beta }^{A\rightarrow AA}|_{AT}=T_{\mu \alpha \beta
}^{A\rightarrow VV}|_{AT}=\frac{-2\pi ^{2}}{(2\pi )^{4}}(b-a)\varepsilon
_{\mu \alpha \beta \lambda }(p-p^{\prime })_{\lambda },
\end{equation}
and finally 
\begin{equation}
T_{\mu \alpha \beta }^{V\rightarrow VV}|_{AT}=T_{\mu \alpha \beta
}^{V\rightarrow AA}|_{AT}=\frac{-2i\pi ^{2}}{3(2\pi )^{4}}(b-a)\left\{
g_{\mu \alpha }(p-p^{\prime })_{\beta }+g_{\mu \beta }(p-p^{\prime
})_{\alpha }+g_{\alpha \beta }(p-q)_{\mu }\right\} .
\end{equation}
The above demonstrations show us, as enounced, that our calculational
technique produces results mappeable in others specific procedures. Let us
now verify the Ward Identities involved.

\section{Ward Identities}

We next proceed to the verification of the free fermion model symmetry
relations presented by eqs.(12)-(47). Such relations were obtained formally
by contracting the Lorentz index of a vertex with the respective external
momentum. There are in principle two ways to verify Ward Identities
involving n-point functions. The first is to contract the integral
representation, before evaluating traces, generating in this way a function
with the same number of points as the contracted one more two others with a
number of points lowered by one unity. Only the former ones are in fact
evaluated explicitly \cite{ref19}. This is the strategy used in ref.\cite
{ref1} to analyze Ward Identities. The second way involves the explicitly
evaluation of the n-point function in question and \underline{afterwards}
contracting the results with the appropriate external momentum, identifying
then all others functions present in the investigated Ward Identities.
Obviously the effort associated is much higher in the second option once all
functions involved needs to be evaluated explicitly. We will analyze in this
work both possibilities for two-point functions and only the first one for
the three-point functions.

\subsection{One-Point Functions}

The first Ward Identity to which we refer in section II involves $T_{\mu
}^{V}$. A vanishing value for this amplitude is predicted in QED by Furry's
theorem, for example. From the result produced by our treatment, eq.(59), we
see that in order to obtain a definite zero value for the $T_{\mu }^{V}$
amplitude in such a way that the result is independent of the arbitrary
choices for the internal momentum the following conditions need to be
satisfied: 
\begin{equation}
\left\{ 
\begin{array}{ll}
\Box _{\alpha \beta \mu \nu }=0 &  \\ 
\nabla _{\mu \nu }=0 &  \\ 
\triangle _{\mu \nu }=0. & 
\end{array}
\right.  \label{eqcon}
\end{equation}
Which we will denominate {\it consistency conditions} \cite{ref24} from now
on. Since all terms of $T_{\mu }^{V}$ in eq.(59) are ambiguous, requiring
relations eqs.(\ref{eqcon}) the amplitude turns unambiguous and symmetry
preserving. As will be shown in what follows they are necessary and
sufficient conditions to ``save'' all other amplitudes from ambiguities and
symmetry violations.

\subsection{Two-Point Functions}

We now study the Ward Identities for the two-point functions. Let us,
initially, consider the contraction of $T_{\mu }^{VS}$ with the external
momentum $(k_{1}-k_{2})_{\mu }$ 
\begin{equation}
(k_{1}-k_{2})_{\mu }T_{\mu }^{VS}=\int \frac{d^{4}k}{(2\pi )^{4}}Tr\left\{ 
\hat{1}\frac{1}{[\not{k}+\not{k}_{2}-m]}({\not{k}}_{1}-{\not{k}}_{2})\frac{1%
}{[\not{k}+\not{k}_{1}-m]}\right\} .
\end{equation}

Before calculating the trace we use the following identity 
\begin{equation}
(\not{k}_{1}-\not{k}_{2})=[\not{k}+\not{k}_{1}-m]-[\not{k}+\not{k}_{2}-m],
\end{equation}
so that 
\begin{equation}
(k_{1}-k_{2})_{\mu }T_{\mu }^{VS}=\int \frac{d^{4}k}{(2\pi )^{4}}Tr\left\{ 
\hat{1}\frac{1}{[\not{k}+\not{k}_{2}-m]}\right\} -\int \frac{d^{4}k}{(2\pi
)^{4}}Tr\left\{ \hat{1}\frac{1}{[\not{k}+\not{k}_{1}-m]}\right\} .
\end{equation}
Comparing now with eq.(60) we identify two scalar one-point functions. It is
clear that the vector current will only be conserved if there is no
dependence on $k_{1}$ and $k_{2}$ in this one-point functions. However,
using the result in eq.(62) for $T^{S}$ we get 
\begin{equation}
(k_{1}-k_{2})_{\mu }T_{\mu }^{VS}=-2m\left[ (k_{1}+k_{2})_{\alpha
}(k_{1}-k_{2})_{\beta }+(k_{1}-k_{2})_{\alpha }(k_{1}+k_{2})_{\beta }\right]
\triangle _{\alpha \beta }.
\end{equation}
We can, on the other hand, also check the validity of the identity by the
contraction of $(k_{1}-k_{2})_{\mu }$ with the expression obtained for $%
T_{\mu }^{VS}$, eq.(80):

\begin{equation}
(k_1-k_2)_\mu T^{VS}_{\mu}=-4m(k_{1}-k_{2})_\mu (k_1+k_2)_\beta
[\triangle_{\mu\beta}].
\end{equation}

In any case the identity can be satisfied provided $\triangle _{\mu \beta
}=0 $, which renders $T^{S}$ unambiguous and $T_{\mu }^{VS}$ zero, as one
can check from eq.(62) and eq.(80). Let us now study the case of
axial-pseudoscalar amplitude $T_{\mu }^{AP}$, defined in eq.(72). Firstly we
do the following 
\begin{equation}
(k_{1}-k_{2})_{\mu }T_{\mu }^{AP}=\int \frac{d^{4}k}{(2\pi )^{4}}Tr\left\{
\gamma _{5}\frac{1}{[\not{k}+\not{k}_{2}-m]}i(\not{k}_{1}-\not{k}_{2})\gamma
_{5}\frac{1}{[\not{k}+\not{k}_{1}-m]}\right\} ,
\end{equation}
and introduce the identity 
\begin{equation}
(\not{k}_{2}-\not{k}_{1})\gamma _{5}=[\not{k}+\not{k}_{2}-m]\gamma
_{5}+\gamma _{5}[\not{k}+\not{k}_{1}-m]+2m\gamma _{5}
\end{equation}
in the interior of the trace, which leads us to 
\begin{eqnarray}
(k_{1}-k_{2})_{\mu }T_{\mu }^{AP} &=&-2miT^{PP}-i\int \frac{d^{4}k}{(2\pi
)^{4}}Tr\left\{ \hat{1}\frac{1}{[\not{k}+\not{k}_{1}-m]}\right\}  \nonumber
\\
&&-i\int \frac{d^{4}k}{(2\pi )^{4}}Tr\left\{ \hat{1}\frac{1}{[\not{k}+\not{k}%
_{2}-m]}\right\} .
\end{eqnarray}
Substituting the results previously obtained for the one-point function we
obtain 
\begin{equation}
(k_{1}-k_{2})_{\mu }T_{\mu }^{AP}=-2miT^{PP}-2mi\left\{
4I_{quad}(m^{2})+2[k_{1\alpha }k_{1\beta }+k_{2\alpha }k_{2\beta }]\triangle
_{\alpha \beta }\right\} .
\end{equation}
Had we, on the other hand, taken eq.(72) and contracting it with $%
(k_{1}-k_{2})_{\mu }$, we would have gotten 
\begin{equation}
(k_{1}-k_{2})_{\mu }T_{\mu }^{AP}=-4mi(k_{1}-k_{2})^{2}\left\{
I_{log}(m^{2})-\left( \frac{i}{(4\pi )^{2}}\right)
Z_{0}(m^{2},m^{2},(k_{1}-k_{2})^{2};m^{2})\right\} ,
\end{equation}
from which we could get 
\begin{eqnarray}
(k_{1}-k_{2})_{\mu }T_{\mu }^{AP} &=&-2mi\left\{
-4I_{quad}(m^{2})+2(k_{1}-k_{2})^{2}I_{log}(m^{2})\right.  \nonumber \\
&&-\left. \left( \frac{i}{(4\pi )^{2}}\right)
(k_{1}-k_{2})^{2}2Z_{0}(m^{2},m^{2},(k_{1}-k_{2})^{2};m^{2})\right. 
\nonumber \\
&&\left. -2(k_{1\alpha }k_{1\beta }+k_{2\alpha }k_{2\beta })\triangle
_{\alpha \beta }\right\}  \nonumber \\
&&-2mi\left\{ 4I_{quad}(m^{2})+2(k_{1\alpha }k_{1\beta }+k_{2\alpha
}k_{2\beta })\triangle _{\alpha \beta }\right\}
\end{eqnarray}
This results agrees with eq.(132) provided we identify, in the first term,
in curly brackets, the expression for $T^{PP}$, as we can see from eq.(71).

Let us next take the amplitude $T^{AV}_{\mu\nu}$ which has two Lorentz
indices and has two Ward Identities connected to it, one for the vector
index and other for the axial one. For the vector current we have 
\begin{equation}
(k_1-k_2)_\nu T^{AV}_{\mu\nu}= \int \frac{d^4k}{(2\pi )^4}Tr \left\{
i\gamma_\mu \gamma_5\frac{1}{[\not{k} +\not{k}_2 -m]}\right\} -\int \frac{%
d^4k}{(2\pi )^4}Tr \left\{ i\gamma_\mu\gamma_5 \frac{1}{[\not{k} +\not{k}_1
-m]}\right\},
\end{equation}
where we immediately identify the one point axial functions which vanish
identically and therefore $(k_1-k_2)_\nu T^{AV}_{\mu\nu}=0$. Doing the same
for the axial current we have

\begin{eqnarray}
(k_1-k_2)_\mu T^{AV}_{\mu\nu}&=&-2mi[T^{PV}_\nu]+ \int \frac{d^4k}{(2\pi )^4}%
Tr \left\{ i\gamma_\nu \gamma_5\frac{1}{[\not{k} +\not{k}_2 -m]}\right\} 
\nonumber \\
& &-\int \frac{d^4k}{(2\pi )^4}Tr \left\{ i\gamma_\nu\gamma_5 \frac{1}{[\not{%
k} +\not{k}_1 -m]}\right\},
\end{eqnarray}
so that 
\begin{equation}
(k_1-k_2)_\mu T^{AV}_{\mu\nu}=-2mi[T^{PV}_\nu],
\end{equation}
yielding the expected result for the Ward Identity in eq.(17). However the
amplitude $T^{PV}_\nu$ is also identically zero given traces properties,
which immediately implies with $(k_1-k_2)_\mu T^{AV}_{\mu\nu}=0$. If we take
the four-divergence directly in eq.(83) it is easy to see that $%
(k_1-k_2)_\nu T^{AV}_{\mu\nu}=0$ and $(k_1-k_2)_\mu T^{AV}_{\mu\nu}=0$. We
will analyze these results in the next section.

The case of the amplitude $T_{\mu \nu }^{VV}$ is analogous. If we first
perform the contraction 
\begin{eqnarray}
(k_{1}-k_{2})_{\mu }T_{\mu \nu }^{VV} &=&\int \frac{d^{4}k}{(2\pi )^{4}}%
Tr\left\{ \gamma _{\nu }\frac{1}{[\not{k}+\not{k}_{2}-m]}\right\}  \nonumber
\\
&&-\int \frac{d^{4}k}{(2\pi )^{4}}Tr\left\{ \gamma _{\nu }\frac{1}{[\not{k}+%
\not{k}_{1}-m]}\right\} ,
\end{eqnarray}
we again see that the result will depend on the $T_{\mu }^{V}$ one-point
functions. Substituting now their expressions we get 
\begin{eqnarray}
(k_{1}-k_{2})_{\mu }T_{\mu \nu }^{VV} &=&4\left\{ (k_{1}-k_{2})_{\alpha
}\nabla _{\alpha \nu }+(k_{1\alpha }k_{1\beta }k_{1\rho }-k_{2\alpha
}k_{2\beta }k_{2\rho })\frac{\Box _{\alpha \beta \rho \nu }}{3}\right. 
\nonumber \\
&&\;\;\;\;\;\left. -(k_{1}^{2}k_{1\rho }-k_{2}^{2}k_{2\rho })\frac{\triangle
_{\rho \nu }}{3}-(k_{1\nu }k_{1\alpha }k_{1\beta }-k_{2\nu }k_{2\alpha
}k_{2\beta })\frac{2}{3}\triangle _{\alpha \beta }\right\}
\end{eqnarray}
The conservation of the vector current demands, therefore that the r.h.s. be
identically null. It is easy to check that similar results would have been
obtained if we had taken the four-divergence directly on the final result
for $T_{\mu \nu }^{VV}$, eq.(91). This shows that in order to satisfy the
Ward Identities relative to the $T_{\mu \nu }^{VV}$ amplitude we need to
require the same conditions as for the one-point function, eq.(123).

Last we turn to $T_{\mu \nu }^{AA}$. Associated to this Green's function we
have two axial currents. This allows us to relate them to the amplitudes $%
T_{\mu }^{AP}$ and $T^{PP}$ by successive contractions with the external
momentum. Thus, upon contracting with $(k_{1}-k_{2})_{\mu }$ we get 
\begin{eqnarray}
(k_{1}-k_{2})_{\mu }T_{\mu \nu }^{AA} &=&-2mi[T_{\nu }^{PA}]+\int \frac{%
d^{4}k}{(2\pi )^{4}}Tr\left\{ \gamma _{\nu }\frac{1}{[\not{k}+\not{k}_{1}-m]}%
\right\}  \nonumber \\
&&-\int \frac{d^{4}k}{(2\pi )^{4}}Tr\left\{ \gamma _{\nu }\frac{1}{[\not{k}+%
\not{k}_{2}-m]}\right\} ,
\end{eqnarray}
where the two last terms are again vector one-point functions and we get 
\begin{eqnarray}
(k_{1}-k_{2})_{\mu }T_{\mu \nu }^{AA} &=&4\left\{ (k_{2}-k_{1})_{\alpha
}\nabla _{\alpha \nu }+(k_{1\alpha }k_{1\beta }k_{1\rho }-k_{2\alpha
}k_{2\beta }k_{2\rho })\frac{\Box _{\alpha \beta \rho \nu }}{3}\right. 
\nonumber \\
&&\;\;\;\;\;\;\left. +(k_{1}^{2}k_{1\rho }-k_{2}^{2}k_{2\rho })\frac{%
\triangle _{\rho \nu }}{3}+(k_{1\nu }k_{1\alpha }k_{1\beta }-k_{2\nu
}k_{2\alpha }k_{2\beta })\frac{2}{3}\triangle _{\alpha \beta }\right\}
-2miT_{\nu }^{PA}.
\end{eqnarray}
Similar results emerges if the contraction is performed with the expression
in eq.(95). Therefore it is verified that the conditions under which the
Ward Identities are satisfied are the same as the previous ones. In order to
obtain the identity from eq.(20) it is enough to contract, once more, with
the index $\nu $. However nothing new is produced.

\subsection{Three-Point Functions}

We next turn to the question of the verification of Ward Identities
involving three-point functions. We will use only the first way pointed out
in the introduction of this section, i.e., by relating the contracted
functions with two-point functions, following Gertsein and Jackiw \cite{ref1}%
.

Let us consider the identity for the amplitude $T_{\lambda }^{VSS}$.
Contracting the $\lambda $ vector Lorentz index with the external momentum
of the vertex; 
\begin{equation}
(k_{3}-k_{2})_{\lambda }T_{\lambda }^{VSS}=\int \frac{d^{4}k}{(2\pi )^{4}}%
Tr\left\{ \hat{1}\frac{1}{[\not{k}+\not{k}_{1}-m]}\hat{1}\frac{1}{[\not{k}+%
\not{k}_{2}-m]}(\not{k}_{3}-\not{k}_{2})\frac{1}{[\not{k}+\not{k}_{3}-m]}%
\right\} ,
\end{equation}
which can be written after the use of the identity 
\begin{equation}
(\not{k}_{3}-\not{k}_{2})=(\not{k}+\not{k}_{3}-m)-(\not{k}+\not{k}_{2}-m)
\end{equation}
in the form: 
\begin{eqnarray}
(k_{3}-k_{2})_{\lambda }T_{\lambda }^{VSS} &=&\int \frac{d^{4}k}{(2\pi )^{4}}%
Tr\left\{ \hat{1}\frac{1}{[\not{k}+\not{k}_{1}-m]}\hat{1}\frac{1}{[\not{k}+%
\not{k}_{2}-m]}\right\}  \nonumber \\
&&-\int \frac{d^{4}k}{(2\pi )^{4}}Tr\left\{ \hat{1}\frac{1}{[\not{k}+\not{k}%
_{1}-m]}\hat{1}\frac{1}{[\not{k}+\not{k}_{3}-m]}\right\} ,
\end{eqnarray}
where one can identify scalar two-point functions i.e., 
\begin{equation}
(k_{3}-k_{2})_{\lambda }T_{\lambda
}^{VSS}(k_{1},m;k_{2},m;k_{3},m)=T^{SS}(k_{1},m;k_{2},m)-T^{SS}(k_{1},m;k_{3},m).
\end{equation}
Looking at eq.(70) we verify that the amplitude $T^{SS}$ possesses
unambiguous terms. In the above expression this means terms which depends on
differences $(k_{1}-k_{2})^{2}$ and $(k_{1}-k_{3})^{2}$ which are the
external momenta $p^{^{\prime }2}$ and $p^{2}$, respectively. Explicitly the
difference in eq.(145) can be written as 
\begin{eqnarray}
&&\!\!\!\!\!\!\!\!(k_{3}-k_{2})_{\lambda }T_{\lambda
}^{VSS}(k_{1},m;k_{2},m;k_{3},m)=  \nonumber \\
&&2\left\{ (4m^{2}-p^{^{\prime }2})\left[ I_{log}(m^{2})-\left( \frac{i}{%
(4\pi )^{2}}\right) Z_{0}(m^{2},m^{2},p^{^{\prime }2};m^{2})\right] \right. 
\nonumber \\
&&-\left. (4m^{2}-p^{2})\left[ I_{log}(m^{2})-\left( \frac{i}{(4\pi )^{2}}%
\right) Z_{0}(m^{2},m^{2},p^{2};m^{2})\right] \right\}  \nonumber \\
&&+2(k_{2\alpha }k_{2\beta }-k_{3\alpha }k_{3\beta })\triangle _{\alpha
\beta }.
\end{eqnarray}
Now we need to include the crossed channel. Redefining adequately the
momenta of the external lines and operating in an analogous way we finally
arrive at 
\begin{eqnarray}
&&\!\!\!\!\!\!\!\!(l_{3}-l_{2})_{\lambda }T_{\lambda
}^{VSS}(l_{1},m;l_{2},m;l_{3},m)=  \nonumber \\
&&2\left\{ (4m^{2}-p^{2})\left[ I_{log}(m^{2})-\left( \frac{i}{(4\pi )^{2}}%
\right) Z_{0}(m^{2},m^{2},p^{2};m^{2})\right] \right.  \nonumber \\
&&-\left. (4m^{2}-p^{^{\prime }2})\left[ I_{log}(m^{2})-\left( \frac{i}{%
(4\pi )^{2}}\right) Z_{0}(m^{2},m^{2},p^{^{\prime }2};m^{2})\right] \right\}
\nonumber \\
&&+2(l_{2\alpha }l_{2\beta }-l_{3\alpha }l_{3\beta })\triangle _{\alpha
\beta }.
\end{eqnarray}
The sum of the contributions of the two channels yield the following
expression for the searched Ward Identity 
\begin{equation}
q_{\lambda }T_{\lambda }^{V\rightarrow SS}=2(k_{2\alpha }k_{2\beta
}-k_{3\alpha }k_{3\beta })\triangle _{\alpha \beta }+2(l_{2\alpha }l_{2\beta
}-l_{3\alpha }l_{3\beta })\triangle _{\alpha \beta }.
\end{equation}
This result shows that the conservation of the vector current depends in
this case on the difference $\triangle _{\alpha \beta }$.

For the amplitude $T^{V\rightarrow PP}_{\lambda}$ we find, proceeding in a
complete similar way as the previous case,

\begin{equation}
q_{\lambda }T_{\lambda }^{V\rightarrow PP}=2(k_{3\alpha }k_{3\beta
}-k_{2\alpha }k_{2\beta })\triangle _{\alpha \beta }+2(l_{3\alpha }l_{3\beta
}-l_{2\alpha }l_{2\beta })\triangle _{\alpha \beta }.
\end{equation}
Still with one Lorentz index we have the process $A\rightarrow SP$ for which
we write first

\begin{equation}
(k_{3}-k_{2})_{\lambda }T_{\lambda }^{ASP}=\int \frac{d^{4}k}{(2\pi )^{4}}%
Tr\left\{ \hat{1}\frac{1}{[\not{k}+\not{k}_{1}-m]}\gamma _{5}\frac{1}{[\not{k%
}+\not{k}_{2}-m]}i(\not{k}_{3}-\not{k}_{2})\gamma _{5}\frac{1}{[\not{k}+\not{%
k}_{3}-m]}\right\} .
\end{equation}
Now, making use of a convenient identity 
\begin{equation}
(\not{k}_{2}-\not{k}_{3})\gamma _{5}=\gamma _{5}[\not{k}+\not{k}_{3}-m]+[%
\not{k}+\not{k}_{2}-m]\gamma _{5}+2m\gamma _{5},
\end{equation}
we can identify on the r.h.s. the two-point functions $T^{PP}$ and $T^{SS}$,
besides $T^{PSP}$. Substituting the results for them with inclusion of the
crossed channel, we have 
\begin{equation}
q_{\lambda }T_{\lambda }^{A\rightarrow SP}=-2mi[T^{P\rightarrow
SP}]-2i(k_{3\alpha }k_{3\beta }-k_{2\alpha }k_{2\beta })\triangle _{\alpha
\beta }-2i(l_{3\alpha }l_{3\beta }-l_{2\alpha }l_{2\beta })\triangle
_{\alpha \beta },
\end{equation}
where the symmetry violating terms are again associated with ambiguities.

The cases with two Lorentz indices are now in order. Making use of the
two-point functions $T_{\mu }^{VS}$, for $T_{\mu \nu }^{SVV}$ we get for the
two associated Ward Identities 
\begin{eqnarray}
p_{\mu }T_{\mu \nu }^{S\rightarrow VV} &=&4m(k_{3}-k_{1})_{\alpha }\triangle
_{\alpha \nu }+4m(l_{1}-l_{2})_{\alpha }\triangle _{\alpha \nu }  \nonumber
\\
&=&8mp_{\alpha }\triangle _{\alpha \nu }
\end{eqnarray}
and 
\begin{eqnarray}
p_{\nu }^{\prime }T_{\mu \nu }^{S\rightarrow VV} &=&4m(k_{1}-k_{2})_{\alpha
}\triangle _{\alpha \mu }+4m(l_{3}-l_{1})_{\alpha }\triangle _{\alpha \mu } 
\nonumber \\
&=&8mp^{\prime }{}_{\alpha }\triangle _{\alpha \mu }.
\end{eqnarray}
In this results it is important to note an unambiguous character for the
violating term. This means that the identity depends on the value of $%
\triangle _{\alpha \mu }$ and this is not associated with ambiguities here.

For the $S\rightarrow AA$ process, which has two related axial currents we
get 
\begin{equation}
p_\mu T^{S\rightarrow AA}_{\mu\nu}=2mi[T^{S\rightarrow PA}_\nu]
-4m(k_2+k_3)_\alpha \triangle_{\alpha\nu}- 4m(l_2+l_3)_\alpha
\triangle_{\alpha\nu}
\end{equation}
and 
\begin{equation}
p^{\prime}_\nu T^{S\rightarrow AA}_{\mu\nu}=2mi[T^{S\rightarrow AP}_\mu]
-4m(k_3+k_2)_\alpha \triangle_{\alpha\mu}- 4m(l_3+l_2)_\alpha
\triangle_{\alpha\mu},
\end{equation}
which can be, in principle, violated by $\triangle_{\mu\nu}$ with ambiguous
coefficients.

Let us now look to the process $V\rightarrow AP$. In the case of the vector
current, the identity is expressed in terms of the two-point functions $%
T_{\mu }^{AP}$, which are unambiguous. Upon inclusion of the crossed channel
we promptly obtain

\begin{equation}
q_{\lambda }T_{\lambda \mu }^{V\rightarrow AP}=0.
\end{equation}
On the other hand, the identity related to the axial current yields 
\begin{equation}
p_{\mu }T_{\lambda \mu }^{V\rightarrow AP}=2mi[T_{\lambda }^{V\rightarrow
PP}]-4m(k_{2}+k_{3})_{\alpha }\triangle _{\alpha \lambda
}-4m(l_{2}+l_{3})_{\alpha }\triangle _{\alpha \lambda }.
\end{equation}
The two identities related to $T_{\mu \nu }^{PVV}$ are satisfied without
restrictions since the two-point functions involved vanish identically. Thus

\begin{equation}
(k_{3}-k_{1})_{\mu }T_{\mu \nu }^{PVV}=0
\end{equation}
and 
\begin{equation}
(k_{1}-k_{2})_{\nu }T_{\mu \nu }^{PVV}=0.
\end{equation}
Also for the processes $V\rightarrow AS$ and $P\rightarrow AA$ both Ward
Identities are satisfied by the same reason. Thus 
\begin{equation}
p_{\mu }T_{\mu \nu }^{S\rightarrow AV}=2miT_{\nu }^{S\rightarrow PV}
\end{equation}
and 
\begin{equation}
p_{\nu }^{\prime }T_{\mu \nu }^{S\rightarrow AV}=0.
\end{equation}
Also 
\begin{equation}
\left\{ 
\begin{array}{c}
p_{\mu }T_{\mu \nu }^{P\rightarrow AA}=2miT_{\nu }^{P\rightarrow PA} \\ 
p_{\nu }^{\prime }T_{\mu \nu }^{P\rightarrow AA}=2miT_{\mu }^{P\rightarrow
AP}.
\end{array}
\right.
\end{equation}

We are now left with the amplitudes with three Lorentz indices. In the case
of $T_{\lambda \mu \nu }^{VVV}$ the identities are expressed in terms of the
two-point functions $T_{\mu \nu }^{VV}$ and we obtain

\begin{eqnarray}
q_\lambda T^{V\rightarrow VV}_{\lambda\mu\nu}&=&\left[(k_1+k_2)_\alpha
(k_1+k_2)_{\beta}-(k_1+k_3)_\alpha(k_1+k_3)_{\beta}\right] \left[%
\Box_{\alpha\beta\mu\nu}-g_{\mu\alpha}\triangle_{\nu\beta} -g_{\nu\alpha}
\triangle_{\mu\beta}-3g_{\mu\nu}\triangle_{\alpha \beta}\right]  \nonumber \\
& & +\left[(l_1+l_2)_\alpha(l_{1}+ l_{2})_\beta-(l_1+l_3)_\alpha(l_{1}+
l_{3})_\beta \right]\left[\Box_{\alpha\beta\mu\nu}-g_{\mu\alpha}\triangle_{%
\nu\beta} -g_{\nu\alpha} \triangle_{\mu\beta}-3g_{\mu\nu}\triangle_{\alpha
\beta}\right]  \nonumber \\
& &+\frac{2}{3}\left[k_{1\beta}
(k_3-k_2)_{\alpha}-k_{1\alpha}(k_3-k_2)_{\beta}
+l_{1\beta}(l_{3}-l_{2})_\alpha -l_{1\alpha}(l_{3}-l_{2})_\beta \right] %
\left[\Box_{\alpha\beta\mu\nu}+g_{\mu\alpha}\triangle_{\nu\beta}
+g_{\nu\alpha} \triangle_{\mu\beta}\right].
\end{eqnarray}

Similar expression for $p_{\mu }T_{\lambda \mu \nu }^{V\rightarrow VV}$ and $%
p_{\nu }^{\prime }T_{\lambda \mu \nu }^{V\rightarrow VV}$ can be obtained.
All conditions are expressed in terms of $\triangle _{\lambda \mu }$ and $%
\Box _{\alpha \beta \nu \lambda }$.In the case of the $T_{\lambda \mu \nu
}^{VAA}$ amplitude we have analogous results. For the vector current we
find, after the standard operations, 
\begin{equation}
(k_{2}-k_{3})_{\lambda }T_{\lambda \mu \nu }^{V\rightarrow
AA}(k_{1},m;k_{2},m;k_{3},m)=T_{\mu \nu }^{AA}(k_{1},m;k_{2},m)-T_{\mu \nu
}^{AA}(k_{1},m;k_{3},m)
\end{equation}
and in consequence of the eq.(95), we may identify 
\begin{equation}
q_{\lambda }T_{\lambda \mu \nu }^{V\rightarrow AA}=-q_{\lambda }T_{\lambda
\mu \nu }^{V\rightarrow VV}.
\end{equation}

For the axial currents, in other hand, we set first 
\begin{equation}
(k_{3}-k_{1})_{\mu }T_{\lambda \mu \nu }^{V\rightarrow AA}=2miT_{\lambda \nu
}^{V\rightarrow PA}+T_{\lambda \nu }^{AA}(k_{1},m;k_{2},m)-T_{\lambda \nu
}^{VV}(k_{3},m;k_{2},m)
\end{equation}
using now the results eq.(91) and eq.(95) and adding the contribution of the
crossed channel, we set 
\begin{eqnarray}
p_{\mu }T_{\lambda \mu \nu }^{V\rightarrow AA} &=&2miT_{\lambda \nu
}^{V\rightarrow PA}  \nonumber \\
&&+\left[ (l_{1}+l_{3})_{\alpha }(l_{1}+l_{3})_{\beta
}-(l_{2}+l_{3})_{\alpha }(l_{2}+l_{3})_{\beta }\right] \left[ \Box _{\alpha
\beta \lambda \nu }-g_{\lambda \beta }\triangle _{\nu \alpha }-g_{\lambda
\alpha }\triangle _{\nu \beta }-3g_{\lambda \nu }\triangle _{\alpha \beta }%
\right]  \nonumber \\
&&-\left[ (k_{1}+k_{2})_{\alpha }(k_{1}+k_{2})_{\beta
}+(k_{2}+k_{3})_{\alpha }(k_{2}+k_{3})_{\beta }\right] \left[ \Box _{\alpha
\beta \lambda \nu }-g_{\lambda \beta }\triangle _{\nu \alpha }-g_{\lambda
\alpha }\triangle _{\nu \beta }-3g_{\lambda \nu }\triangle _{\alpha \beta }%
\right]  \nonumber \\
&&+\frac{2}{3}\left[ k_{2\alpha }(k_{1}+k_{3})_{\beta }-k_{2\beta
}(k_{1}+k_{2})_{\alpha }+l_{3\alpha }(l_{1}-l_{2})_{\beta }-l_{3\beta
}(l_{1}-l_{2})_{\alpha }\right] \left[ \Box _{\alpha \beta \lambda \nu
}+g_{\nu \alpha }\triangle _{\lambda \beta }+g_{\lambda \alpha }\triangle
_{\nu \beta }\right] .
\end{eqnarray}
A completely analogous condition can be obtained for the other axial current
involved in $T_{\lambda \mu \nu }^{V\rightarrow AA}$.

Now we can finally investigate those associated to anomalous Ward Identity.
First, for the $AVV$ amplitude, contracting with $(k_3-k_2)_\lambda$ before
taking the traces, and using the identity 
\begin{equation}
(\not{k}_2-\not{k}_3)\gamma_5=(\not{k}+\not{k}_2-m)\gamma_5+ \gamma_5(\not{k}%
+\not{k}_3-m)+2m\gamma_5
\end{equation}
we get 
\begin{eqnarray}
(k_3-k_2)_\lambda T^{AVV}_{\lambda\mu\nu}&=& -2mi\int \frac{d^4k}{(2\pi )^4}%
Tr \left\{ \gamma_\mu \frac{1}{[\not{k} +\not{k}_1 -m]} \gamma_\nu\frac{1}{[%
\not{k} +\not{k} _2-m]} \gamma_5\frac{1}{[\not{k} +\not{k} _3-m]}\right\} 
\nonumber \\
& &-\int \frac{d^4k}{(2\pi )^4}Tr \left\{ i\gamma_\nu \gamma_5 \frac{1}{[%
\not{k} +\not{k}_3 -m]} \gamma_\mu \frac{1}{[\not{k} +\not{k}_1-m]}\right\} 
\nonumber \\
& &+\int \frac{d^4k}{(2\pi )^4}Tr \left\{ i\gamma_\mu\gamma_5 \frac{1}{[\not{%
k} +\not{k}_1 -m]} \gamma_\nu \frac{1}{[\not{k} +\not{k}_2-m]}\right\},
\end{eqnarray}
which is the usual procedure \cite{ref1}, \cite{ref4}, \cite{ref5}, \cite
{ref6}.

On the right hand side we can easily identify the three-point function $PVV$
and a pair of two-point functions $AV$. Substituting the appropriate value
for two-point functions involved, eq.(83), and adding the corresponding
crossed channel we obtain for the Ward Identity

\begin{eqnarray}
q_{\lambda }T_{\lambda \mu \nu }^{A\rightarrow VV} &=&-2mi[T_{\mu \nu
}^{P\rightarrow VV}]  \nonumber \\
&&+2\varepsilon _{\mu \nu \alpha \beta }\left[ (k_{1}-k_{3})_{\beta
}(k_{1}+k_{3})_{\xi }+(k_{2}-k_{1})_{\beta }(k_{1}+k_{2})_{\xi }\right]
\triangle _{\xi \alpha }  \nonumber \\
&&-2\varepsilon _{\mu \nu \alpha \beta }\left[ (l_{1}-l_{3})_{\beta
}(l_{1}+l_{3})_{\xi }+(l_{2}-l_{1})_{\beta }(l_{1}+l_{2})_{\xi }\right]
\triangle _{\xi \alpha }
\end{eqnarray}
For the vector currents, first we contract with $(k_{3}-k_{1})_{\mu }$
momentum, the direct channel, and make use of the identity

\begin{equation}
(\not{k}_3-\not{k}_1)=(\not{k}+\not{k}_3-m)-(\not{k}+\not{k}_1-m),
\end{equation}
to obtain the relation

\begin{eqnarray}
(k_{3}-k_{1})_{\mu }T_{\lambda \mu \nu }^{AVV} &=&\int \frac{d^{4}k}{(2\pi
)^{4}}Tr\left\{ i\gamma _{\lambda }\gamma _{5}\frac{1}{[\not{k}+\not{k}%
_{1}-m]}\gamma _{\nu }\frac{1}{[\not{k}+\not{k}_{2}-m]}\right\}  \nonumber \\
&&-\int \frac{d^{4}k}{(2\pi )^{4}}Tr\left\{ i\gamma _{\lambda }\gamma _{5}%
\frac{1}{[\not{k}+\not{k}_{3}-m]}\gamma _{\nu }\frac{1}{[\not{k}+\not{k}%
_{2}-m]}\right\} .
\end{eqnarray}
again the value assumed by $AV$ two-point function is crucial to define the
value of the right hand side of the above equation. Using the results
obtained in our calculation we get, after the addition of the crossed
channel results,

\begin{eqnarray}
p_\mu T^{A\rightarrow VV}_{\lambda\mu\nu} &=&
+2\varepsilon_{\lambda\nu\alpha\beta}\left[ (k_{2}-k_{1})_%
\beta(k_{1}+k_{2})_\xi + (k_{3}-k_{2})_\beta(k_{2}+k_{3})_\xi \right]%
\triangle_{\xi\alpha}  \nonumber \\
& &+2\varepsilon_{\lambda\nu\alpha\beta}\left[ (l_{3}-l_{1})_%
\beta(l_{1}+l_{3})_\xi + (l_{2}-l_{3})_\beta(l_{2}+l_{3})_\xi \right]%
\triangle_{\xi\alpha}.
\end{eqnarray}
In a completely analogous way, we arrive at the expression, for the other
vector current involved;

\begin{eqnarray}
p^{\prime}_\nu T^{A\rightarrow VV}_{\lambda\mu\nu} &=&
+2\varepsilon_{\lambda\mu\alpha\beta}\left[ (k_{3}-k_{1})_%
\beta(k_{1}+k_{3})_\xi + (k_{2}-k_{3})_\beta(k_{2}+k_{3})_\xi \right]%
\triangle_{\xi\alpha}  \nonumber \\
& &+2\varepsilon_{\lambda\mu\alpha\beta}\left[ (l_{2}-l_{1})_%
\beta(l_{1}+l_{2})_\xi + (l_{3}-l_{2})_\beta(l_{2}+l_{3})_\xi \right]%
\triangle_{\xi\alpha}.
\end{eqnarray}

We repeat then the procedure for the three Ward Identities associated to $%
T_{\lambda \mu \nu }^{AAA}$. In this case, the contraction with the external
momenta give rise again to two-point functions $T_{\mu \nu }^{AV}$ which,
after some manipulation, yields

\begin{eqnarray}
q_{\lambda }T_{\lambda \mu \nu }^{A\rightarrow AA} &=&-2miT_{\mu \nu
}^{P\rightarrow AA}  \nonumber \\
&&+2\varepsilon _{\mu \nu \alpha \beta }\left[ (k_{3}-k_{1})_{\beta
}(k_{1}+k_{3})_{\xi }+(k_{1}-k_{2})_{\beta }(k_{1}+k_{2})_{\xi }\right]
\triangle _{\xi \alpha }  \nonumber \\
&&+2\varepsilon _{\mu \nu \alpha \beta }\left[ (l_{3}-l_{1})_{\beta
}(l_{1}+l_{3})_{\xi }+(l_{1}-l_{2})_{\beta }(l_{1}+l_{2})_{\xi }\right]
\triangle _{\xi \alpha }
\end{eqnarray}
and, in a completely analogous fashion, 
\begin{eqnarray}
p_{\mu }T_{\lambda \mu \nu }^{A\rightarrow AA} &=&2miT_{\lambda \nu
}^{A\rightarrow PA}  \nonumber \\
&&-2\varepsilon _{\lambda \nu \alpha \beta }\left[ (k_{1}-k_{2})_{\beta
}(k_{1}+k_{2})_{\xi }+(k_{3}-k_{2})_{\beta }(k_{2}+k_{3})_{\xi }\right]
\triangle _{\xi \alpha }  \nonumber \\
&&+2\varepsilon _{\lambda \nu \alpha \beta }\left[ (l_{3}-l_{2})_{\beta
}(l_{1}+l_{3})_{\xi }+(l_{1}-l_{3})_{\beta }(l_{1}+l_{3})_{\xi }\right]
\triangle _{\xi \alpha }.
\end{eqnarray}
Finally, for the last Ward Identity, we get: 
\begin{eqnarray}
p_{\nu }^{\prime }T_{\lambda \mu \nu }^{A\rightarrow AA} &=&2miT_{\lambda
\mu }^{A\rightarrow AP}  \nonumber \\
&&+2\varepsilon _{\lambda \mu \alpha \beta }\left[ (k_{1}-k_{3})_{\beta
}(k_{1}+k_{3})_{\xi }+(k_{3}-k_{2})_{\beta }(k_{2}+k_{3})_{\xi }\right]
\triangle _{\xi \alpha }  \nonumber \\
&&+2\varepsilon _{\lambda \mu \alpha \beta }\left[ (l_{3}-l_{2})_{\beta
}(l_{2}+l_{3})_{\xi }+(l_{1}-l_{2})_{\beta }(l_{1}+l_{2})_{\xi }\right]
\triangle _{\xi \alpha }.
\end{eqnarray}

Let us now proceed the final analysis of the preceding investigations.

\section{Final Analysis}

For a free $1/2$ fermion, with equal masses, we have established the
symmetry relations involving the divergent one, two and three-point
functions. We next explicitly calculated, with arbitrary choices for the
internal momenta of the loops, one and two-point functions and evaluated the
potentially ambiguous part of the three-point functions. All the
calculations have been performed from the point of view of a general
calculational method in which only general assumptions are made in respect
of an eventual regulator taken in an implicit way. No explicit calculations
of a divergent integral are performed. In the result for the specific
divergent integral, once no shifts are assumed, the results corresponding to
other techniques are still maintained. To map these results into specific
procedures all we need is to interpretate or evaluate the divergent terms
according to the desired philosophy. In particular we have shown in the
section V that the results corresponding to the traditional reference on
this subject are still present in our results and a complete mapping is
immediate. We are then at the position to analyze all the results looking at
them and having in mind a searching for an universal interpretation of the
possibilities of ambiguities and symmetry relation violations. In particular
we are worried to answer the question: the justification of the anomaly
phenomena can be consistently given in terms of ambiguities associated with
(physically irrelevant) choices for the internal momenta (breaking of
translational invariance)?. Our main argument to expect a negative answer to
this question resides on the fact that if the anomalies are fundamental
quantum phenomena therefore they must not be conditioned or justified in
terms of divergences aspects, since, if the exact solutions were possible,
then the description of the dynamics of the interacting particle would not
involve divergences and, consequently, ambiguities. In other words, we
believe that the ambiguities are typical consequences of the perturbative
solution approach, as well as the infinities are, and a consistent
interpretation of the perturbative amplitudes should leave no room for
ambiguities like the renormalization process needs to eliminate all the
divergences to furnish physical amplitudes. Any consistent treatment of the
mathematical indefinitions associated to divergent integrals, from the
physical point of view, needs to automatically eliminate the ambiguities.
Let us now come from what we believe to what we obtain strictly from the
point of view of the mathematical treatment.

We first note that {\it in all calculated amplitudes the ambiguous
dependence on the internal arbitrary momenta are always coefficients of only
three combinations of divergent objects}, with the same degree of
divergence. They are

\begin{eqnarray}
\Box_{\alpha \beta \mu \nu }&=&\int_\Lambda \frac{d^4k}{(2\pi )^4}\frac{
24k_\mu k_\nu k_\alpha k_\beta }{(k^2 -m^2 )^4} -g_{\alpha \beta}
\int_\Lambda \frac{d^4k}{(2\pi )^4}\frac{4k_\mu k_\nu }{[(k^2 -m^2)^3]} 
\nonumber \\
& &-g_{\alpha \mu} \int_\Lambda \frac{d^4k}{(2\pi )^4}\frac{4k_\beta k_\nu }{%
[(k^2 -m^2)^3]} -g_{\alpha \nu} \int_\Lambda \frac{d^4k}{(2\pi )^4}\frac{%
4k_\beta k_\mu }{[(k^2 -m^2)^3]}, \\
\nabla_{\mu\nu}&=&\int_\Lambda \frac{d^4k}{(2\pi )^4}\frac{2k_\mu k_\nu }{%
(k^2 -m^2 )^2}-\int_\Lambda \frac{d^4k}{(2\pi )^4}\frac{g_{\mu\nu}}{(k^2
-m^2 )}, \\
\triangle_{\mu\nu}&=&\int_\Lambda\frac{d^4k}{(2\pi )^4}\frac{4k_\mu k_\nu }{%
(k^2 -m^2 )^3}-\int_\Lambda \frac{d^4k}{(2\pi )^4}\frac{g_{\mu\nu}}{(k^2
-m^2 )^2}.
\end{eqnarray}

The second crucial observation is that {\it all violating terms are also
combinations of these objects but not all of them possess ambiguous
coefficients}. This means that even if shifts were allowed without
restrictions the violating terms would be not completely removed. This
statement is deeply related to the success of the Dimensional Regularization
technique, which is well-known for producing amplitudes in a consistent way
in respect ambiguities and symmetry relations. It is very easy to verify
that in DR the objects $\Box $, $\nabla $ and $\triangle $ vanish
identically in such a way that all ambiguous and symmetry violating terms
are automatically removed. Note that it is not sufficient to authorize
shifts but it is necessary that the properties between divergent integrals,
which we denominate consistency conditions, are assumed by the method
because these aspects are not always associated. So, to obtain a perfect map
from our result to those corresponding to DR calculations it is sufficient
to eliminate $\Box $, $\nabla $ and $\triangle $ and to evaluate the objects 
$I_{log}(m^{2})$ and $I_{quad}(m^{2})$ from the point of view of that
technique (taking the apropriate value for the traces of the $\gamma $
matrices $tr(\gamma _{\mu }\gamma _{\nu })=2^{\omega }g_{\mu \nu }$ and so
on). The finiteness part is perfectly mapped one-by-one. These
identifications are strictly possible for the amplitudes that can be
obtained in $2\omega $ dimensions ($\omega $ = continuum and complex), i.e.,
to the cases where the number of the $\gamma _{5}$ involved matrix is not an
odd quantity \cite{ref25}. At this point it is not possible to run away from
the observation: from our result it is possible to recover the DR results,
where this technique could be applied, assuming $\Box =\nabla =\triangle =0$%
. It is also possible to recover the surface term evaluations of the
Gertsein-Jackiw approach in which case the result for $\Box $, $\nabla $ and 
$\triangle $ cannot be zero. If we follow the way guided by DR philosophy,
in what concerns the interpretation of the divergent objects, then we have
obtained, from our general approach, all Ward Identity satisfied and all
ambiguities are removed by the same condition that leads to this fact. Then
the triangle anomaly phenomenon cannot be justified in a consistent way
based on the divergent character of the involved amplitudes (ambiguities).
If in the other hand we follow the Gertsein and Jackiw procedure then
violations in Ward Identities of $AVV$ and $AAA$ amplitudes are obtained but
a hard price must be paid because the ambiguities and violations would
propagate to many other amplitudes which is a very undesirable feature of
QFT in general, once renormalization become a very complicated scheme to be
achieved in this fashion.

Another aspect involved in the Gertsein-Jackiw justification of the triangle
anomaly phenomenon, which is quoted in many textbooks of QFT even nowadays 
\cite{ref5}\cite{ref6}, is that the source of the violating term (anomaly)
resides strictly on the value to be assumed by the $AV$ two-point function.
This approach needs a nonzero value for this physical amplitude. This is a
clearly inconsistent statement from many aspects of the QFT features. First,
if the $AV$ amplitude does not vanish identically, it possesses two
conserved currents: an axial-vector and a vector one, once we had: 
\begin{eqnarray}
&&(k_{i}-k_{j})_{\mu }T_{\mu \nu }^{AV}(k_{i},m;k_{j},m)=0  \nonumber \\
&&(k_{i}-k_{j})_{\nu }T_{\mu \nu }^{AV}(k_{i},m;k_{j},m)=0,
\end{eqnarray}
which is not consistent with the symmetry content of the model, in the sense
that, to justify the violation of a symmetry relation we first need to
violate another one, in such a way it is not possible to decide what is the
fundamental symmetry breaking phenomenon. A second aspect is related to the
unitarity. If the $AV$ amplitude does not vanish it needs to develop an
imaginary part when the external momentum is $\left( k_{1}-k_{2}\right)
^{2}=4m^{2}$ (Cutkosky's rules). The value attributed to $AV$ amplitude by
the Gertsein-Jackiw approach is not compatible with this QFT perturbative
aspect. In addition, a CPT breaking may be introduced because the transition
between a vector to axial-vector particle seems to be possible if the $AV$
two-point function becomes a nonzero quantity. Now we return again to
specific analysis of our calculations.

In the present context, the analysis of the results, regarding the
ambiguities and symmetry relations, resulted rather transparent, once we
look all the conditions through the three consistency conditions above
cited. Concerning this two aspects we can state what follows:

\begin{itemize}
\item  {\bf Ambiguities}

In all amplitudes the dependence arbitrary choices of internal momenta of
loops appear simply as coefficients for the differences between divergent
integrals, eqs.(178)-(180). Given this fact the conclusion is immediate: all
the ambiguities will be eliminated if they are simultaneously zero. However,
it is not surprising that ambiguities could be eliminated if shifts were
allowed, i.e., provided we ignore the corresponding surface's terms, as we
can see observing the conditions (123). Conversely, from this point of view,
the conclusion can be looked as 4-D conditions to be satisfied by any
regularization prescription which should have the consistency of DR whenever
it applies. But this is not the whole story.

\item  {\bf Ward Identities}

In our investigations we have found several examples in which ambiguities
and symmetry violations are intimately connected and have the same origin.
However, there were also several instances in which Ward Identities could be
violated by unambiguous terms with the same structure of the relations in
the eqs.(178)-(180). This is the case, for example, of $T_{\mu \nu }^{VV}$
and $T_{\mu \nu }^{AA}$, where the difference $\nabla _{\alpha \beta }$
appears with an unambiguous coefficient. Also the three-point functions $%
T_{\mu \nu }^{SVV}$ and $T_{\mu \nu }^{SAA}$ involve the difference $%
\triangle _{\alpha \beta }$ as a condition for the fulfillment of the
corresponding Ward Identities. The difference $\Box _{\alpha \beta \mu \nu }$
would also appear with an unambiguous coefficient for the four-point
function $T_{\mu \nu \lambda \rho }^{VVVV}$, among others, as it can be
easily checked.
\end{itemize}

We could then invert the analysis starting precisely by these amplitudes and
extracting the conclusion that, independently of ambiguities, the objects $%
\Box_{\alpha \beta \mu \nu }$, $\nabla_{\mu\nu}$ and $\triangle_{\mu\nu}$
should be obtained as zero in a consistent treatment. We would then, a
posteriori, verify that these conditions eliminate all sources of
ambiguities.

At this point we reach an important and quite surprising result: Following
the strategy of Gertsein and Jackiw in ref.\cite{ref1} to study Ward
Identities, which has historically been used to justify violations of
symmetry relations, we found a set of conditions which allows all the WI to
be satisfied. In this context the possibility of making use of ambiguities
for any purpose is automatically eliminated. By imposing these referred
conditions, the corresponding results of DR can be immediately mapped
whenever it can be applied. Those corresponding Gertsein and Jackiw results,
as we have shown in section V, can be equally obtained from our results but
not with the same interpretation for the objects $\Box _{\alpha \beta \mu
\nu }$, $\nabla _{\mu \nu }$ and $\triangle _{\mu \nu }$.

The situation is now the following: To establish or justify the existence of
the anomaly phenomena, in the context of perturbative calculations, we need
to use a specific prescription to evaluate some divergent integrals. The
traditional one, based on surface terms, could be used, in principle, to
treat all the amplitudes for any theories, but it is discarded nowadays
where the DR can be applied and it is accepted only for the treatment of
pseudo-amplitudes where the DR cannot be used due to its natural limitation.
The two treatments lead to results that cannot be mapped one into another
for places where both can be applied. If we are looking for an universal way
to treat all divergent amplitudes in QFT, the above situation is
unacceptable.

We arrive at two deeply different options: first, if we adopt the
interpretation corresponding to the surface terms point of view we can get a
picture for triangle anomalies that corresponds to the one of the Gertsein
an Jackiw, but, in consequence, we will plague all physical amplitudes with
ambiguities and consequently loose the translational invariance, the main
one of the basic space-time symmetries implemented in the construction of
QFT's. Second, if we adopt the DR interpretation for the objects $\Box
_{\alpha \beta \mu \nu }$, $\nabla _{\mu \nu }$ and $\triangle _{\mu \nu }$
we will have all symmetry relations satisfied, including those considered as
anomalous.

This statement is an immediate consequence of our strategy in looking at
perturbative calculations involving divergent amplitudes; the consistency
conditions make immaterial an eventual choice for the value of undefined
quantities, because in all places of occurrence they are multiplied by
differences between divergent integrals of the same degree of divergence,
that need to be identically zero by construction.

Once we have concluded that there is no chance of consistency in
calculations involving divergent integrals without the imposition of the
consistent conditions, a crucial question emerges: Do the combination of the
treatment given to the divergent integral plus the strategy of Gertsein and
Jackiw to verify Ward identities in three-point functions, lead to the
conclusion that there are no anomalies in triangle diagrams? The answer is:
not necessarily. The conclusion of our investigation, which at this point
becomes transparent, is that this kind of analysis is not completely
consistent, because we can find conditions that produce exactly the opposite
conclusion of the initial intentions.

What is then the correct procedure to discuss this problem? Our expectation
resides on the explicit calculations for the three-point functions and
contracting them with the momenta only after the calculations are performed
to generate Ward Identities. The correct violation values for symmetry
relations (anomalies) required by phenomenological reasons like the neutral
pion decay, need to emerge in a natural way, free from ambiguities related
to the arbitrary choices of internal labels. The anomalous term needs to be
associated with intrinsic properties of the involved three-point functions $%
T_{\lambda \mu \nu }^{A\rightarrow VV}$ and $T_{\mu \nu }^{P\rightarrow VV}$
and not with the value for a two-point function, $T_{\mu \nu }^{AV}$, which
needs to be identically zero by many reasons. This is actually the essential
point of the Sutherland-Veltman paradox \cite{ref26} that states the
impossibility to obtain simultaneously the three Ward Identities satisfied
and the correct value for the amplitude $AVV$ in zero axial vertex momentum
by a smooth limit. In the calculations here performed, and many others
cited, these ingredients are not present and it is not reasonable to extract
a conclusion about the triangle anomaly phenomenon.

In fact a work along these lines has been performed and a manuscript where
all details involved in a completely general analytic calculation is in
preparation \cite{ref27}. Some results can be anticipated to give support to
our final expectations and suppositions. After all the necessary
calculations to give the most general analytic expression for $T_{\lambda
\mu \nu }^{A\rightarrow VV}$ have been concluded, we can identify only one
ambiguous term with the coefficient $\Delta _{\alpha \beta }$. The
contraction with the external momenta gives us 
\begin{eqnarray}
\bullet p_{\mu }T_{\lambda \mu \nu }^{A\rightarrow VV} &=&\left( \frac{i}{%
4\pi ^{2}}\right) \varepsilon _{\nu \beta \lambda \xi }p_{\xi }p_{\beta
}^{\prime } \\
\bullet p_{\nu }^{\prime }T_{\lambda \mu \nu }^{A\rightarrow VV} &=&-\left( 
\frac{i}{4\pi ^{2}}\right) \varepsilon _{\mu \beta \lambda \xi }p_{\xi
}p_{\beta }^{\prime } \\
\bullet q_{\lambda }T_{\lambda \mu \nu }^{A\rightarrow VV} &=&-2mi\left\{
T_{\mu \nu }^{P\rightarrow VV}\right\} ,
\end{eqnarray}
which shows that in adopting the value suggested by our investigation, $%
\Delta _{\alpha \beta }=0$, we \ do not spoil the violation. In addition we
note that the value of the violating term is exactly that related to the $%
\lim_{q_{\lambda }\rightarrow 0}\;q^{\lambda }T_{\lambda \mu \nu }^{AVV}$ or
that one we need for the phenomenological adjustment on the $A\rightarrow VV$
amplitude to allow, the neutral pion decay with a correct width. So, in the
context of the Sutherland-Veltman paradox, if we choose an amplitude with a
correct behavior on the $q_{\lambda }\rightarrow 0$ limit and simultaneously 
$U(1)$ gauge invariant for all values of the external momenta then an
anomalous term should be included leading to the Ward identities 
\begin{eqnarray}
\bullet p_{\nu }^{\prime }\left( T_{\lambda \mu \nu }^{A\rightarrow
VV}\right) _{phy} &=&0 \\
\bullet p_{\mu }\left( T_{\lambda \mu \nu }^{A\rightarrow VV}\right) _{phy}
&=&0 \\
\bullet q_{\lambda }\left( T_{\lambda \mu \nu }^{A\rightarrow VV}\right)
_{phy} &=&-2mi\left\{ T_{\mu \nu }^{P\rightarrow VV}\right\} -\left( \frac{i%
}{2\pi ^{2}}\right) \varepsilon _{\mu \nu \alpha \beta }p_{\alpha }p_{\beta
}^{\prime }.
\end{eqnarray}
Then, the $PVV$ amplitude in the right hand side of the eq. (188) can be
identified as responsible for the neutral pion decay. It is important to
emphasize the main aspect involved; the resulting amplitudes were derived
without having recourse to ambiguities. The space-time homogeneity has been
maintained. In addition the source of the anomalous term resides in strictly
finite terms and then it is not necessarily associated to properties of
divergent integrals and neither to the specific aspects of perturbative
calculations. This is precisely what we should expect of a fundamental
quantum phenomenon: to be still present in the eventual exact solutions
consequently not revealed to us by the nature throughout the infinities
which are nothing more than a consequence of our inability in solving the
equations of motion of a QFT in an exactly way. Finally we call the
attention that our results are in complete agreement with what we need to
construct the renormalization of the Standard Model by anomaly cancellation.

\end{document}